\documentclass[article]{revtex4}

\usepackage{latexsym,amssymb,float}
\usepackage{setspace}
\usepackage{graphicx}
\usepackage{epsfig}
\usepackage{amsmath}
\usepackage{bm}

\begin{document}

\title{Theory of Scanning Tunneling Spectroscopy: from Kondo Impurities to Heavy Fermion Materials}
\author{Dirk K. Morr}
\affiliation{Department of Physics, University of Illinois at
Chicago, Chicago, IL 60607, USA}
\date{\today}

\begin{abstract}

\end{abstract}

\begin{abstract}
Kondo systems ranging from the single Kondo impurity to heavy fermion materials present us with a plethora of unconventional properties whose
theoretical understanding is still one of the major open problems in condensed matter physics. Over the last few years, groundbreaking scanning tunneling spectroscopy (STS) experiments have provided unprecedented new insight into the electronic structure of Kondo systems. Interpreting the results of these experiments -- the differential conductance and the quasi-particle interference spectrum -- however, has been complicated by the fact that electrons tunneling from the STS tip into the system can tunnel either into the heavy magnetic moment or the light conduction band states. In this article, we briefly review the theoretical progress made in understanding how quantum interference between these two tunneling paths affects the experimental STS results. We show how this theoretical insight has allowed us to interpret the results of STS experiments on a series of heavy fermion materials providing detailed knowledge of their complex electronic structure. It is this knowledge that is a {\it conditio sine qua non} for developing a deeper understanding of the fascinating properties exhibited by heavy fermion materials, ranging from unconventional superconductivity to non-Fermi-liquid behavior in the vicinity of quantum critical points.

\end{abstract}

\maketitle

\section{Introduction}

\begin{figure}[h]
%
%
\begin{center}
\includegraphics[height=4.5cm]{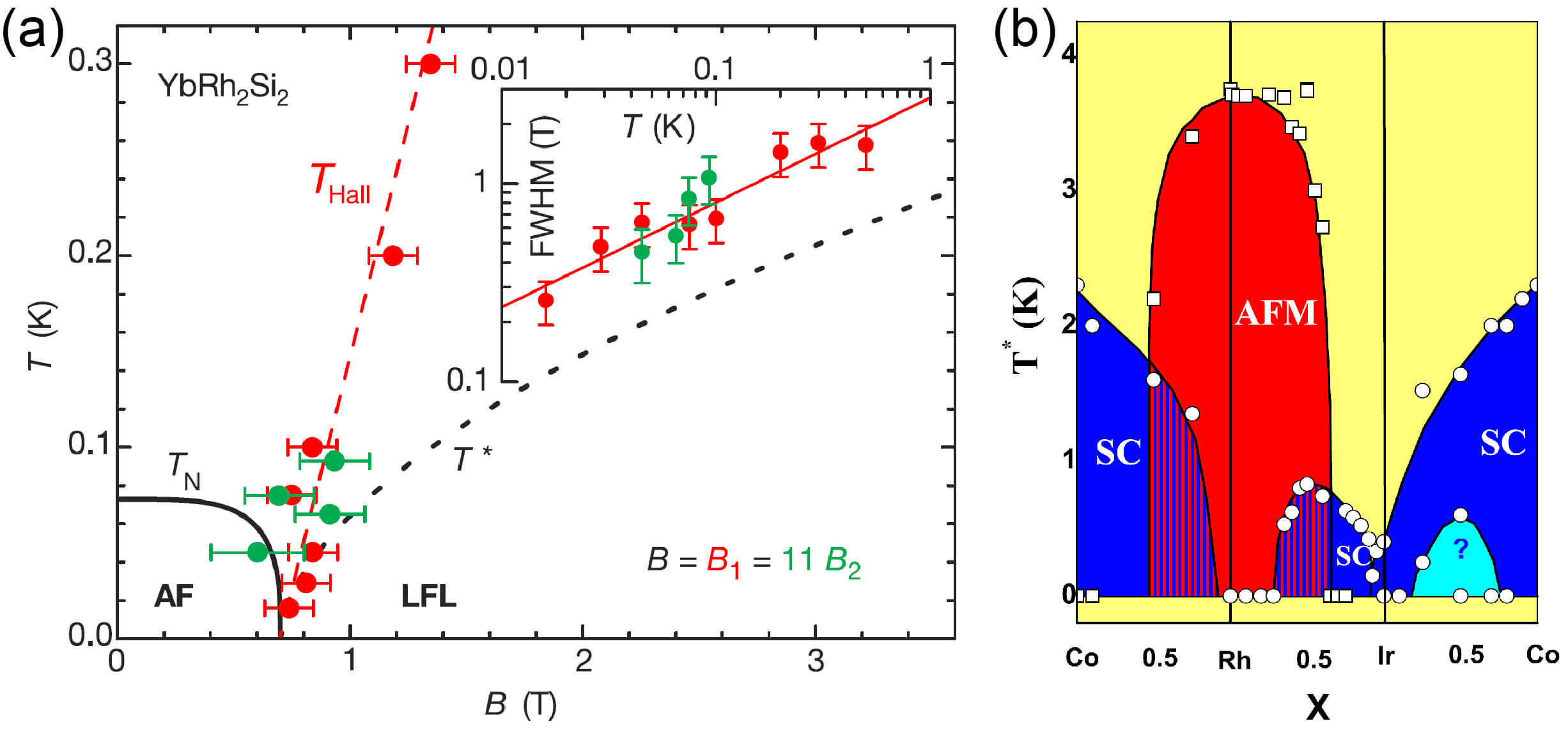}
\caption{Phase diagrams of (a) YbRh$_2$Si$_2$ \cite{Pas04}, and (b) the ``115" compounds \cite{Sar07}.}
\label{fig:PD}
\end{center}
%
%
\end{figure}

The study of the Kondo effect \cite{Kondo64} from the single magnetic impurity to heavy fermion materials \cite{Don77}, has remained one of the most fascinating topics in condensed matter physics since its discovery more than 80 years ago \cite{deH34}. One of the key unresolved challenges in this field is to identify the microscopic mechanism giving rise to the complex phase diagram of heavy fermion materials, and their many unconventional properties \cite{Ste01,Mon07,Loh07,Si10,Nor11,Ste12,Sca12}. The most salient features of their phase diagrams are an antiferromagnetically long-range ordered phase with an associated magnetic quantum critical point (QCP) \cite{ASch00,Cus03,Pas04,Geg05,Col07}, and a Kondo screened, heavy-Fermi-liquid region, as shown in Fig.~\ref{fig:PD}(a) for the prototypical heavy fermion material YbRh$_2$Si$_2$ \cite{Cus03,Pas04}. The great interest in heavy fermion materials arises from two intriguing phenomena associated with this QCP. Some heavy fermion materials, such as YbRh$_2$Si$_2$, possess properties \cite{Ste01,Loh07,Loh94,Map95,Tro00} in the quantum critical region \cite{Col05b,Geg08} above the QCP which violate the predictions of Landau's Fermi liquid theory -- one of the cornerstones of modern condensed matter physics -- and hence are labelled non-Fermi liquid (NFL) properties. Other heavy fermion materials, such as the ``115" compounds \cite{Pet01,Mos01,Iza01,Sar07,Sto08}, exhibit unconventional superconducting phases close to the QCP \cite{Ste79,Mat98,Sid02,Sto11} [see Fig.~\ref{fig:PD}(b)].
To-date, no consensus has emerged on the microscopic origin of either of these two phenomena. While it is generally believed that the unconventional superconducting phase arises from $f$-electron magnetism \cite{Mon07,Nor11,Sca12,Ste12}, the lack of detailed insight into the momentum structure of the heavy bands and of the superconducting gap has made the unambiguous identification of the pairing mechanism all but impossible. The same lack of insight into the form of the electronic and magnetic excitations in the quantum critical region has also hindered a deeper understanding of the observed NFL properties which have been attributed  to the presence of massless spin fluctuations \cite{Mor95,Ros97,Ros99,Abr12}, the competition between the antiferromagnetically ordered and Kondo screened phases \cite{Don77,Si01,Ing02,Sun03,Gre06,Mar10,Tan11,Col01,Col99,Sen03,Sen04}, critical fluctuations of the hybridization \cite{Paul07}, and disorder effects \cite{Neto98,Mil02,Sch11,Mir05}. Identifying the key aspects responsible for the complex properties of heavy-fermion materials, is therefore one of the major open problems in condensed matter physics.

A major breakthrough in resolving this important problem has recently been achieved by scanning tunneling spectroscopy (STS) experiments \cite{Sch09,Ayn10,Ern11,Ham11,Wahl11,Ayn12,All13,Zhou13,Ena15} investigating the complex electronic structure of a series of important heavy fermion materials in their normal and superconducting states.  These experiments measure the spatially and energy resolved differential conductance, $dI/dV$, which in materials with a single electronic band is proportional to the local density of states (LDOS). An important technique employed in these experiments is quasi-particle interference (QPI) spectroscopy \cite{Sch09,Ayn12,All13,Zhou13}. Its main idea is that the spatial oscillations induced by defects in the differential conductance, $dI/dV$, are dominantly $2k_F r$ oscillations, arising from the backscattering of electrons across the Fermi surface, and hence leading to a change of $2k_F$ in the electrons momentum. By fourier-transforming these real space oscillations into momentum space, one can therefore in principle map out the electronic structure of a material, as has been successfully demonstrated not only in simple metals \cite{Crom93,Has93} but also in the cuprate \cite{Hoff02} and iron-based superconductors \cite{Han10}.

By measuring both the differential conductance as well as the QPI spectrum, STS experiments have investigated the electronic structure of a  series of intriguing heavy fermion materials: (i) URu$_2$Si$_2$ \cite{Sch09,Ayn10}, which undergoes a puzzling second order phase transition at $T_0 =17.5$K
into a state with a still unknown, {\it hidden order parameter} \cite{Pal85,Map86,Scho87,Bonn88,Dor01,Rod97,Esc94,Den01,Cox87,Cha02,Var06,Bal09,Elg09,Pep11,Hau09,Opp10,Dubi11,Cha13,Myd14,Oka11,Kung15}, (ii) YRh$_2$Si$_2$ \cite{Ern11}, whose phase diagram exhibits a magnetic quantum critical point, and (iii) CeCoIn$_5$ \cite{All13,Zhou13}, a material considered to be the ``hydrogen" atom for our understanding of unconventional superconductivity in heavy fermion materials.
\begin{figure}[h]
\includegraphics[height=5.cm]{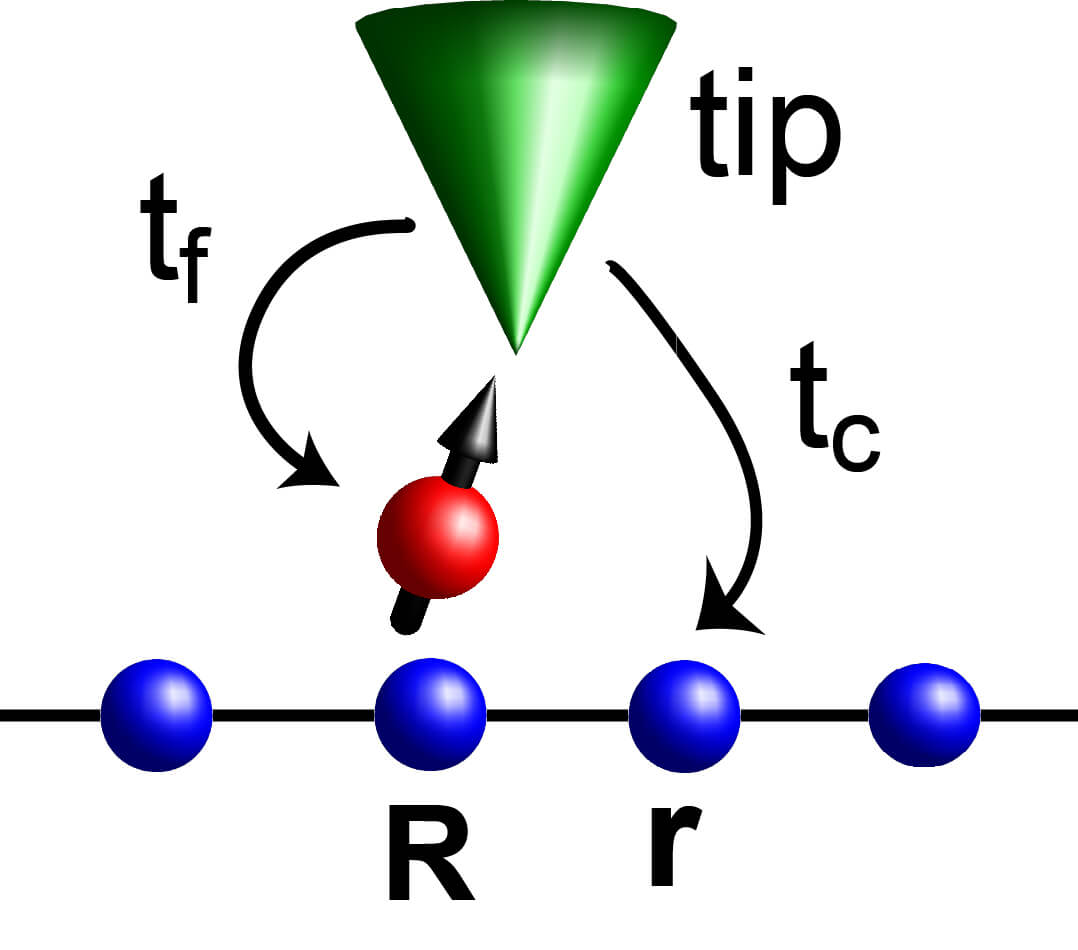}
\caption{Tunneling paths of electrons from the STS tip
into the conduction and $f$-electron states with tunneling amplitudes $t_c$
and $t_f$, respectively \cite{Fig10}. The filled red and blue circles represent the magnetic atom and the surface atoms, respectively.} \label{fig:Fig1}
\end{figure}
While these experiments might hold the key to understanding the complex properties of these materials, a difficulty in interpreting the experimental results arises from identifying the relation between the measured differential conductance or the QPI spectrum, and the electronic structure of heavy fermion materials \cite{Fano61}. In particular,
quantum interference between electrons tunneling from the STS tip into the conduction band and into the states containing the magnetic moment (see Fig.~\ref{fig:Fig1}), has rendered the interpretation of $dI/dV$, even for the case of single magnetic defects on metallic surfaces, quite difficult. For this reason, the $dI/dV$ data taken near isolated magnetic defects \cite{Mad98,Li98,Ujs00,Kno02,Ali05,Mad01} were often interpreted using a phenomenological expression first derived by Fano \cite{Fano61}.

However, motivated by the experimental breakthroughs in performing STS experiments on heavy fermion materials, a series of theoretical studies \cite{Mal09,Fig10,Wol10,Fig11,Yuan12,Par13,Dyke14,Pet14} have recently emerged that have provided  a microscopic understanding of how the interplay between the strength of the Kondo coupling, the interaction between the magnetic moments, the electronic structure of the screening conduction band, and quantum interference determines the $dI/dV$ lineshape. These studies have also extracted the detailed momentum structure of the complex, hybridized electronic bands \cite{Yuan12}, and identified the symmetry and momentum dependence of the superconducting gaps \cite{All13}. This in turn has enabled the development of a quantitative understanding of the microscopic mechanism underlying the emergence of unconventional superconductivity in heavy fermion materials \cite{Dyke14}.

The rest of the paper is organized as follows. In Sec.~\ref{sec:theory} we review the theoretical formalism that establishes the relation between the differential conductance and the QPI spectrum measured in STS experiments, and the electronic structure of a single magnetic defect (Sec.~\ref{sec:TheorySingleDefect}), and of heavy fermion materials (Secs.~\ref{sec:AndMod1} and \ref{sec:AndMod2}). In Sec.~\ref{sec:QI_SingleKondo} we discuss the experimental $dI/dV$ lineshapes around isolated magnetic defects, and demonstrate how they are determined by quantum interference effects. In Sec.~\ref{sec:HF} we review STS experiments on URu$_2$Si$_2$ (Sec.~\ref{sec:URuSi}) and CeCoIn$_5$  (Sec.~\ref{sec:CeCoIn5}), and the novel insight they provided into the electronic structure of heavy fermion materials. In Sec.~\ref{sec:Defects}, we discuss how defects in heavy fermion materials affect their electronic structure, and give rise to hybridization waves. Finally, in Sec.~\ref{sec:Concl} we present our conclusions and provide an outlook on current and future work.

\section{Formalism}
\label{sec:theory}

Quantum interference in Kondo systems and heavy fermion materials is directly tied to their multi-orbital or multi-band character. In the following, we briefly outline how the differential conductance can be computed in the presence of multiple tunneling paths, and how the quantum interference between these paths determines the $dI/dV$ lineshape, and the corresponding quasi-particle interference spectrum.

\subsection{Tunneling into a single Kondo impurity}
\label{sec:TheorySingleDefect}

To demonstrate the importance of quantum interference in determining the lineshape of the differential conductance, $dI/dV$,  we begin by considering a system with a single magnetic impurity located on a metallic surface (see Fig.~\ref{fig:Fig1}), described by the Kondo Hamiltonian \cite{Fig10}
\begin{equation}
{\cal H} = \sum_{{\bf k},\sigma} \varepsilon_{\bf k}
c^\dagger_{{\bf k},\sigma} c_{{\bf k},\sigma} + J {\bf S}_{\bf
R} \cdot {\bf s}^c_{\bf R} \ , \label{eq:1}
\end{equation}
where $\varepsilon_{\bf k}$ is the conduction band dispersion, and
$c^\dagger_{{\bf k},\sigma}$ ($c_{{\bf k},\sigma}$) creates
(annihilates) a conduction electron with spin $\sigma$ and momentum ${\bf k}$. ${\bf
S}_{\bf R}$ and ${\bf s}^c_{\bf R}$ are the spin operators of
the magnetic impurity and the conduction electrons at site ${\bf R}$,
respectively, and $J>0$ is the Kondo coupling. To describe the Kondo screening of the magnetic impurity, we use a fermionic $SU(N)$ representation of the spin operators \cite{Coq69,Read83,Col83,Bic87,Mil87} via
\begin{equation}
{\bf S}_{\bf r} =  \sum_{\alpha, \beta} f^\dagger_{{\bf
r},\alpha} {\bm \Gamma}_{\alpha, \beta} f_{{\bf r},\beta} \ ; \quad
{\bf s}^{c}_{\bf r} =  \sum_{\alpha, \beta} c^\dagger_{{\bf
r},\alpha} {\bm \Gamma}_{\alpha, \beta} c_{{\bf r},\beta} \ ,
 \label{eq:slave} 
\end{equation}
where ${\bf \Gamma} = (\Gamma^1, ...,\Gamma^M)$ are the $M=N^2-1$ independent generators of $SU(N)$ in the fundamental representation, $N=2S+1$ is the spin degeneracy of the magnetic moment, and $f^\dagger_{{\bf r},\alpha}$, $f_{{\bf r},\alpha}$ are the Abrikosov pseudofermion operators that represent the magnetic moment. The pseudofermion operators are subject to the constraint
$n_f=\sum_{\alpha=1..N} f^\dagger_{{\bf r},\alpha} f_{{\bf r},\alpha}=1$. Within a path integral approach, this constraint is
enforced by means of a Lagrange multiplier $\varepsilon_f$, while
the exchange interaction in Eq.(\ref{eq:1}) is decoupled using a Hubbard Stratonovich transformation and introducing the hybridization field $s$. Here, a non-zero hybridization implies screening of the magnetic moment. By minimizing the effective action on the (static) saddle point level, we obtain two self-consistent equations (considering the case $S=1/2$ and hence $N=2$) given by
\begin{subequations}
\begin{align}
s &= -\frac{J}{\pi} \int_{ - \infty}^{\infty} d\omega \ n_F(\omega) {\rm Im} G^r_{fc}({\bf R},{\bf R}, \omega) \label{eq:hyb} \ ;\\
n_f &= 1 = -\frac{1}{\pi} \int_{ - \infty}^{\infty} d\omega \ n_F(\omega) {\rm Im} G^r_{ff}({\bf R},{\bf R}, \omega) \ , \label{eq:epsf}
\end{align}
\end{subequations}
where $n_F(\omega)$ is the Fermi distribution function, and $G^r$ is the full retarded Greens function arising from the hybridization process with
\begin{subequations}
\begin{align}
G_{ff}^r({\bf R}, {\bf R}, \omega) & =   \left[\omega + i \delta -\varepsilon_f
- s^2 g^r_0({\bf R}, {\bf R}, \omega)\right] ^{-1} \ ; \label{eq:GF1} \\
G_{cc}^r({\bf r}, {\bf r}, \omega) & =  g^r_0({\bf r}, {\bf r},
\omega) + g^r_0({\bf r}, {\bf R}, \omega) s \, G^r_{ff}({\bf R}, {\bf
R}, \omega) s \, g^r_0({\bf R}, {\bf r}, \omega) \ ;  \\
G^r_{cf}({\bf r}, {\bf R}, \omega) & =  g^r_0({\bf r}, {\bf R},
\omega) s \, G^r_{ff}({\bf R}, {\bf R}, \omega) \ . \label{eq:GF3}
\end{align}
\end{subequations}
Here, $g^r_0$ is the retarded Greens function of the unhybridized
conduction electron band. In Matsubara $\tau$-space, these Green's functions are defined via $G_{\alpha \beta}({\bf r}^\prime, {\bf r},
\tau)=-\langle T_\tau \alpha^\dagger_{{\bf r}^\prime}(\tau)
\beta_{\bf r}(0) \rangle$ ($\alpha,\beta=c,f$). We note that Eqs.(\ref{eq:hyb}) and (\ref{eq:epsf}) are employed to determine the hybridization $s$ and the renormalized energy of the $f$-levels, $\varepsilon_f$. Their solutions, together with the Green's functions of Eqs.(\ref{eq:GF1}) - (\ref{eq:GF3}) fully describe  the
many-body effects arising from the hybridization of the conduction
band with the $f$-electron level, and the concomitant screening of
the magnetic moment.

To compute the differential conductance measured in STS experiments, we note that an electron tunneling from the STS tip into the system can tunnel either into a conduction electron state at ${\bf r}$ or the $f$-electron state at ${\bf R}$, as schematically shown in Fig.~\ref{fig:Fig1}, allowing for the emergence of quantum interference between these tunneling paths. If the STS tip is positioned above a site ${\bf r}$ of the surface, these tunneling processes are described by the Hamiltonian \cite{Fig10,Wol10}
\begin{equation}
{\cal H}_T =  t_f({\bf r-R}) \sum_{\sigma} f^\dagger_{{\bf R},\sigma} d_{\sigma}  + \sum_{{\bf r'},\sigma}  t_c({\bf r-r'}) c^\dagger_{{\bf r'},\sigma} d_{\sigma} +  H.c.
\ , \label{eq:tun}
\end{equation}
where $d_{\sigma}$ destroys an electron with spin $\sigma$ in the STS tip. Here, $t_c({\bf r-r'})$ and $t_f({\bf r-R})$ are the distance dependent tunneling amplitudes  between the tip and a site ${\bf r'}$ on the metallic surface or the site ${\bf R}$ of the magnetic $f$-level, respectively.  For simplicity, it is assumed that the tunneling is ``on-site", i.e., $t_c({\bf r-r'})=t_c \, \delta_{\bf r,r'}$ and $t_f({\bf r-R})=t_f \, \delta_{\bf r,R}$ since tunneling to nearest neighbor sites is strongly suppressed due to the rapid spatial decay of orbital wave-functions involved in the tunneling processes. Moreover, due to the strong Coulomb repulsion in the magnetic $f$-electron level, one expects that the tunneling amplitude $t_f$ is significantly smaller than $t_c$ even when the STS tip is positioned directly above the magnetic atom. We will see that this expectation is borne out by the theoretical analysis of the experimentally measured $dI/dV$ lineshapes.

Assuming that the STS tip is positioned above the magnetic atom at site ${\bf R}$, the total current flowing from the STS tip into the system's conduction band and $f$-level is given by \cite{Car71}
\begin{eqnarray}
I(V)&=&-\frac{e}{\hbar} \, {\rm Re} \, \int_0^{eV} \frac{d \omega}{2
\pi} \left[ t_c \, {\hat {\cal G}}^K_{12}(\omega) + t_f \, {\hat
{\cal  G}}_{13}^K(\omega)  \right] \ ,
 \label{eq:IV}
\end{eqnarray}
where ${\hat {\cal  G}}^K(\omega)$ is the Keldysh Green's function matrix that accounts for the tunneling between the tip and the system, and is given by
\begin{align}
{\hat {\cal G}}^K(\omega) &= [{\hat 1} - {\hat G}^r(\omega) {\hat t}]^{-1}
{\hat F}(\omega) [{\hat 1} -  {\hat t} {\hat G}^a(\omega) ]^{-1}
\end{align}
where
\begin{subequations}
\begin{align}
{\hat F}(\omega) &= 2i \left(1-2 {\hat n_F}(\omega)
\right) {\rm
Im} \left[ {\hat G}^r(\omega) \right] \ ; \\
{\hat G}^r(\omega) &=
\begin{pmatrix} G^r_t(\omega) & 0 & 0 \\ 0 & G^r_{cc}({\bf R}, {\bf
R}, \omega)& G^r_{cf}({\bf R}, {\bf R},
\omega) \\
0 & G^r_{fc}({\bf R}, {\bf R}, \omega) & G^r_{ff}({\bf R}, {\bf R},
\omega)
\end{pmatrix} \ ,
\end{align}
\end{subequations}
and the elements of ${\hat G}^r$ are given in Eqs.(\ref{eq:GF1}) - (\ref{eq:GF3}). Here, ${\hat t} $ is the symmetric tunneling matrix that contains the non-zero tunneling elements between the tip and the system given by
${\hat t}_{12} = t_c$, ${\hat t}_{13} = t_f$. ${\hat n_F}$
is a diagonal matrix containing the Fermi-distribution functions of the tip, the
$f$- and $c$-electron states, and $G^r_t$ is the retarded Greens
function of the tip.

\begin{figure}[h]
\includegraphics[height=3.5cm]{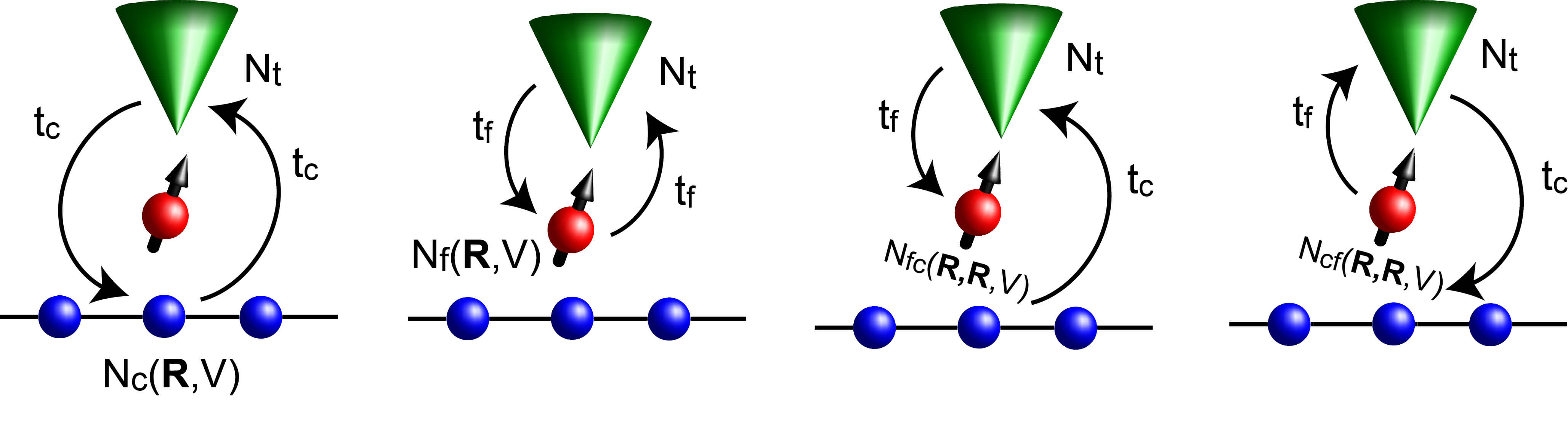}
\caption{Tunneling processes which contribute to the differential conductance [see Eq.(\ref{eq:dIdV})].} \label{fig:TP}
\end{figure}
To gain insight into the physical quantities that govern the flow of current from the tip into the system, and ultimately determine the differential conductance, $dI/dV$,
we consider the experimentally relevant weak-tunneling limit, $t_c,t_f
\rightarrow 0$. In this case, we expand the right hand side of Eq.(\ref{eq:IV}) to leading order in the tunneling elements, thus obtaining for the differential conductance
\begin{align}
\frac{dI(V)}{dV} & = \frac{2 \pi e^2}{\hbar} N_t \left[t_c^2
N_{c}({\bf R}, V) + t_f^2 N_{f}({\bf R}, V) +  t_c t_f N_{cf}({\bf R}, V) + t_f t_c N_{fc}({\bf R}, V) \right] \  ,
\label{eq:dIdV}
\end{align}
where $N_t, N_c$ and $N_f$ are the density of states of the tip, the
conduction and $f$-electron states, respectively, with $N_{c}=-{\rm
Im} G^r_{cc}({\bf R}, {\bf R}, V)/\pi$ and $N_{f}=-{\rm
Im} G^r_{ff}({\bf R}, {\bf R}, V)/\pi$.  Moreover, $N_{cf}=-{\rm
Im} G^r_{cf}({\bf R}, {\bf R}, V)/\pi$ and $N_{fc}=-{\rm
Im} G^r_{fc}({\bf R},{\bf R},  V)/\pi$ represent the correlations between the $f$-state and the conduction electron state at ${\bf R}$ arising from the hybridization. The last two terms yield identical contributions to the differential conductance. All four terms in Eq.(\ref{eq:dIdV}) can be visualized as closed paths on which electrons tunnel from the tip into the system and back, as shown in Fig.\ref{fig:TP}. It is the interference between these four tunneling processes that ultimately determines the $dI/dV$ lineshape, as discussed below.

\subsection{Tunneling into a Heavy Fermion Material}
\label{sec:AndMod1}

Our starting point for the description of heavy fermion materials
is the $U \rightarrow \infty$ limit of the Anderson model \cite{Col84,AL,Kot88,Hew97,Col07} which allows for charge fluctuations in the electronic levels containing the magnetic moments. The corresponding Hamiltonian is given by
\begin{align}
{\cal H} &= \sum_{{\bf k},\sigma} \varepsilon_{\bf k}
c^\dagger_{{\bf k},\sigma} c_{{\bf k},\sigma} + \sum_{{\bf r},\sigma} E_0
f^\dagger_{{\bf r},\sigma} f_{{\bf r},\sigma} - V_0 {\sum_{{\bf r},\sigma}} \left( f^\dagger_{{\bf r},\sigma}
b_{{\bf r}} c_{{\bf r},\sigma} + H.c. \right)
+ {\sum_{{\bf r,r'}}} I_{{\bf r,r'}} {\bf S}_{\bf r} \cdot
{\bf S}_{\bf r'} \label{eq:AndModel} \ ,
\end{align}
where $f^\dagger_{{\bf r},\sigma}$ creates an electron
with spin $\sigma$ at site ${\bf r}$ in the heavy $f$-band, and $V_0$ is the (bare) hybridization between the $c$- and $f$-bands.
To account for valence fluctuations between unoccupied and singly occupied $f$-electron sites,
one introduces the slave-boson operators $b^\dagger_{{\bf r}},b_{{\bf r}}$ and the constraint
$\sum_\sigma f^\dagger_{{\bf r},\sigma} f_{{\bf r},\sigma} + b^\dagger_{{\bf r}}
b_{{\bf r}} = 1 $ which ensures an $f$-electron occupancy $n_f<1$. Moreover, $I_{{\bf r,r'}}$ is the
antiferromagnetic interaction between magnetic moments in the $f$-band. The origin of the magnetic interaction can lie
either in direct exchange or arise from the Ruderman-Kittel-Kasuya-Yosida (RKKY) interaction \cite{Rud54,Kas56,Yos57} mediated by the conduction electrons.
Insight into the complex electronic
bandstructure of heavy fermion materials provided by STS experiment (see Secs.\ref{sec:HF}) have opened new possibilities to identify the origin of the magnetic interaction.

Similar to the single Kondo impurity case, one uses the path integral approach and employs the pseudo-fermion representation for ${\bf S}_{\bf r}$ \cite{Kot88,Read83,Col83,Col84,Hew97} and decouples the magnetic interaction term using a Hubbard-Stratonovich field, $t_f({{\bf r,r'}, \tau})$. The constraint is enforced by means of a Lagrange multiplier $(\epsilon_f-E_0)$. In the static saddle point approximation (and in the radial gauge \cite{Col84,AL})
one replaces $b^\dagger_{{\bf r}}, b_{{\bf r}}$ by their expectation value $\langle b^\dagger_{{\bf r}} \rangle = r_0({\bf r}) e^{i \phi({\bf r})}$ and
subsumes the phase factor $e^{i \phi}$ into a redefinition of the fermionic-operators $f^\dagger,f$. A condensation of the bosonic operators (i.e.,
$r_0 \not = 0$) represents the screening of the magnetic moments. Moreover, the field $t_f({{\bf r,r'}, \tau})$ is replaced by its static expectation value $t_f({\bf r},{\bf r}')$ which describes the antiferromagnetic correlations \cite{Sen03,Paul07,Fig10} between magnetic moments.
Minimizing the effective action, one then obtains the following set of self-consistent equations
\begin{subequations}
\begin{align}
s({\bf r}) &= -\frac{J_0}{\pi} \int_{-\infty}^{\infty} d\omega \
n_F(\omega) \ {\rm
Im} G_{fc}({\bf r},{\bf r},\omega) \ ;   \label{eq:Andsc1} \\
t({{\bf r,r'}}) &= -\frac{I_{\bf r,r'}}{\pi}
\int_{-\infty}^{\infty} d\omega \ n_F(\omega) \ {\rm Im} G_{ff}({\bf
r},{\bf r^\prime},\omega)  \ ; \label{eq:Andsc2}  \\
n_f({\bf r}) &= - 
\int_{-\infty}^{\infty} \frac{d\omega}{\pi} \ n_F(\omega) \ {\rm Im} G_{ff}({\bf
r},{\bf r},\omega) \ , \label{eq:Andsc3}
\end{align}
\end{subequations}
where $n_f({\bf r}) = 1 - r_0^2({\bf r})$, $J_0=V^2/(\varepsilon_f-E_0)>0$, and $s({\bf r})=V_0r_0({\bf r})$ is the effective hybridization with $s({\bf r})=s$ for translationally invariant systems. Note that these self-consistent equations possess the same functional form as those for the Kondo model, Eqs.(\ref{eq:hyb}) and (\ref{eq:epsf}), and that within the mean-field approach described here, the self-consistent equations for the Kondo lattice model are obtained from those of the Anderson lattice model, Eqs.(\ref{eq:Andsc1}) - (\ref{eq:Andsc3}), in the limit $r_0 \rightarrow 0$.   Moreover, if we assume that the magnetic interaction occurs only between nearest and next-nearest-neighbor sites ${\bf r},{\bf r}'$, then for a translationally invariant system, we have $t_f({\bf r},{\bf r}')=t_{f1}$ and $t_{f2}$ for
nearest and next-nearest-neighbor sites, respectively. This yields a dispersion of the heavy $f$-band given by
\begin{equation}
\varepsilon^f_{\bf k} = -2 t_{f1}(\cos k_x+\cos k_y)-4 t_{f2} \cos k_x
\cos k_y + \varepsilon_f \ .
\end{equation}
Moreover, in this mean-field approximation, the full Green's functions in momentum space, which describe the hybridization between the $c$- and $f$-electron bands, are given by
\begin{subequations}
\begin{eqnarray}
G_{ff}({\bf k},\alpha, \omega) & = &  \left[(G_{ff}^0({\bf k}, \alpha, \omega))^{-1} - s^2
G_{cc}^0({\bf k}, \alpha, \omega) \right] ^{-1} \ ;  \\
G_{cc}({\bf k}, \alpha, \omega) & = & \left[(G_{cc}^0({\bf k}, \alpha, \omega))^{-1} - s^2
G_{ff}^0({\bf k}, \alpha, \omega) \right] ^{-1} \ ;  \\
G_{cf}({\bf k}, \alpha, \omega) & = & - G_{cc}^0({\bf k}, \alpha, \omega) s
G_{ff}({\bf k}, \alpha, \omega) \ , \label{eq:GF}
\end{eqnarray}
\end{subequations}
where $G_{ff}^0 = (\omega + i \Gamma_f -\varepsilon^f_{\bf k})^{-1}, G_{cc}^0
= (\omega + i \Gamma_c -\varepsilon^c_{\bf k})^{-1}$, and
$\Gamma^{-1}_c$ and $\Gamma^{-1}_f$ are the lifetimes of the $c$-
and $f$-electron states, respectively. For $\Gamma_c=\Gamma_f=0^+$,
the poles of the above Green's functions yield two energy bands with dispersion
\begin{equation}
E_{\bf k}^{\pm}=\frac{\varepsilon^c_{\bf k} + \varepsilon^f_{\bf k}}{2} \pm
\sqrt{\left( \frac{\varepsilon^c_{\bf k} - \varepsilon^f_{\bf k}}{2}\right)^2 +
s^2} \ .
\label{eq:dispersions}
\end{equation}
For the heavy fermion material URu$_2$Si$_2$, it was argued \cite{Yuan12} that the valence fluctuations occur between singly and doubly occupied $f$-electron sites which leads to $n_f>1$. In order to describe this case, it is necessary to perform a particle-hole
transformation of the slave-boson Anderson Hamiltonian, in which case the constraint takes the form $\sum_\sigma f^\dagger_{{\bf r},\sigma} f_{{\bf r},\sigma} - b^\dagger_{{\bf r}}
b_{{\bf r}} = 1 $ and consequently $n_f({\bf r}) = 1 + r_0^2({\bf r})$. However, the form of the self-consistent equations, Eqs.(\ref{eq:Andsc1}) - (\ref{eq:Andsc3}), remains unchanged.

Finally, we want to briefly mention that the last decade has also seen the development of a series of numerical approaches, such as the dynamical mean-field theory (DMFT) \cite{Sun03,Hau10,Ben11} or the dynamical cluster approach \cite{Mar08,Wu15}, that have been successful in describing various aspects of heavy fermion materials. While these approaches are limited in their ability to describe momentum-resolved properties of these materials, they account for incoherent processes associated with the formation of the heavy Fermi liquid state.

\begin{figure}[h]
\includegraphics[height=3.5cm]{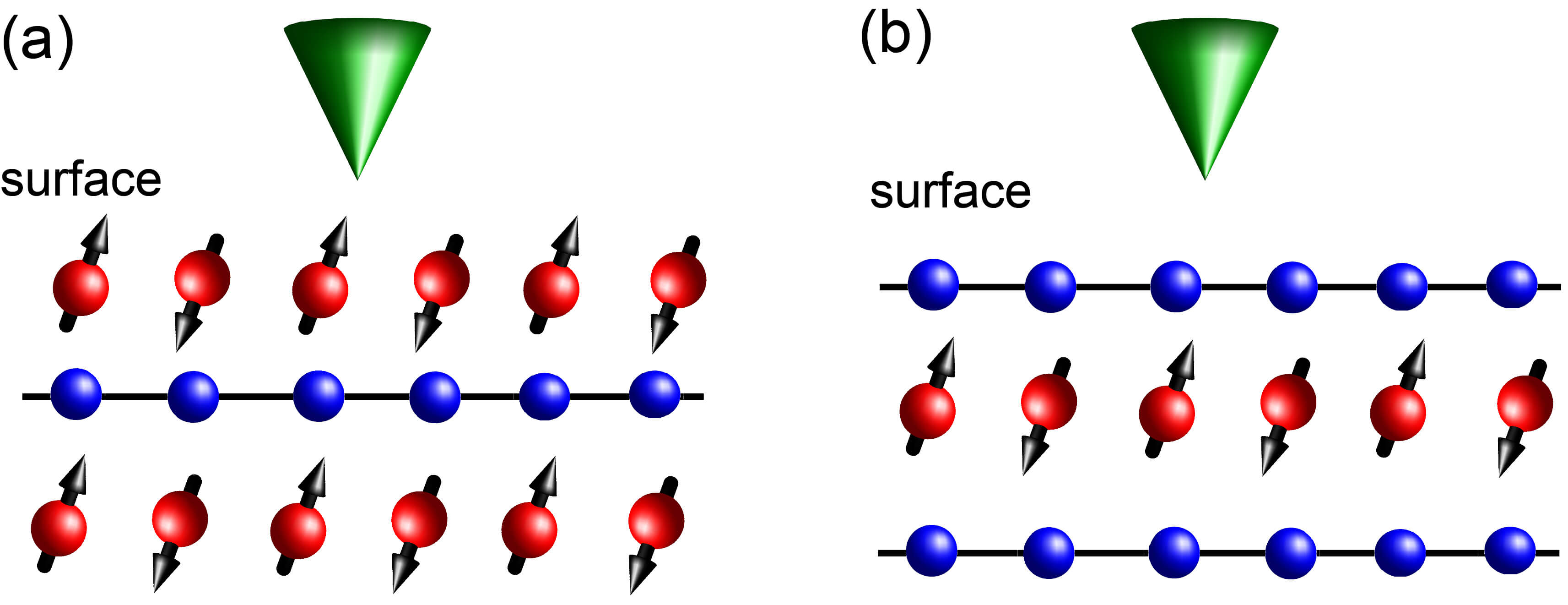}
\caption{The nature of the surface termination layer in heavy fermion materials determines the form of the $dI/dV$ lineshape through $t_f/t_c$. (a) The termination layer consists of $f$-moments, suggesting a larger value of $t_f/t_c$. (b) The termination layer is a conduction band layer, implying a small or vanishing value of $t_f/t_c$. The filled red and blue circles represent the layer of magnetic atoms and of conduction band sites, respectively.} \label{fig:Surface}
\end{figure}

In the Anderson model, the tunneling process into a heavy fermion material is described by the Hamiltonian \cite{Yuan12}
\begin{align}
{\cal H}_T &= \sum_{{\bf r},\sigma} \left[ t_c c^\dagger_{{\bf r},\sigma} d_{\sigma} +
t_f^{(0)} f^\dagger_{{\bf r},\sigma} b_{{\bf r}} d_{\sigma} + H.c. \right]\ , \label{eq:tun_And}
\end{align}
where we again assume ``on-site" tunneling only. Within the saddle-point approximation, the effective tunneling into the $f$-electron states is given by $t_f=t_f^{(0)} r_0$, and the differential conductance is obtained from Eq.(\ref{eq:dIdV}). We expect that the experimental $dI/dV$ lineshapes should sensitively depend on whether the surface termination layer is a layer of $f$-moments [see Fig.~\ref{fig:Surface}(a)], in which case $t_f/t_c$ should be larger and the $dI/dV$ lineshape is dominated by the local electronic structure of the $f$-electrons, or a conduction band layer [see Fig.~\ref{fig:Surface}(b)] in which case one expects $t_f/t_c$ to be small and $dI/dV$ to be determined by the local electronic structure of the $c$-electrons. It was recently suggested that the situation might be even more complicated if the magnetic atoms do not only possess magnetic $f$-levels, but also conduction electron states \cite{Pet14} that directly interact with the magnetic moment via the Kondo coupling. We note in passing that $t_f^{(0)}$ and $t_c$ can in general be computed using first principle methods: this would require not only exact knowledge of the orbitals in the STM tip and the heavy fermion material that are involved in the tunneling process, but also to account for the strong Coulomb repulsion in the magnetic $f$-levels.

\begin{figure}[h]
\includegraphics[height=3.25cm]{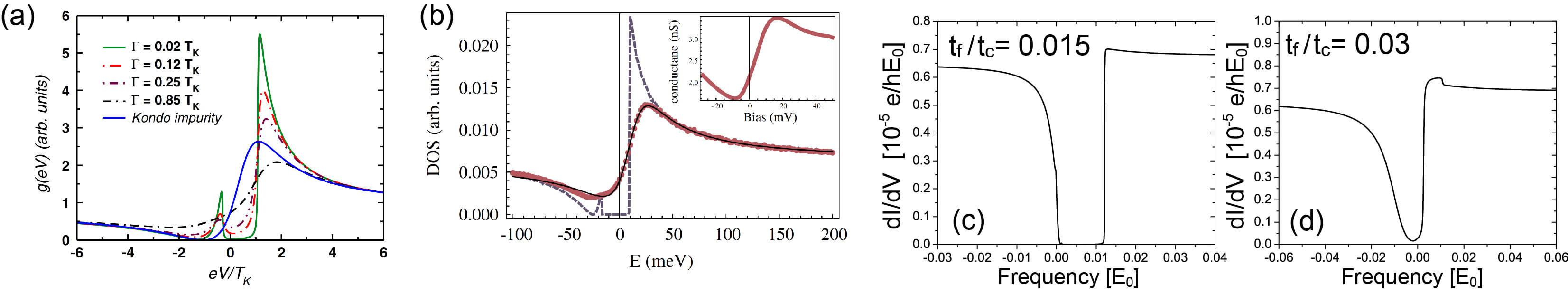}
\caption{(a) Evolution of $dI/dV$ in a Kondo lattice with increasing disorder \cite{Mal09}. (b) $dI/dV$ in a Kondo lattice in the presence of inelastic scattering processes \cite{Wol10}. $dI/dV$ in a Kondo lattice (c) in the presence of an indirect hybridization gap, and (d) when only a direct and no indirect gap is present in the hybridized bandstructure \cite{Fig10}.} \label{fig:Kondo_Lattice}
\end{figure}

Maltseva {\it et al.}\cite{Mal09} and Woelfle {\it et al.}\cite{Wol10} considered a tunneling Hamiltonian similar to that in Eq.(\ref{eq:tun_And}) to investigate the form of the differential conductance in Kondo lattice systems [see Figs.~\ref{fig:Kondo_Lattice}(a) and (b)]
Specifically, Maltseva {\it et al.} proposed that in addition to a direct tunneling process of an electron from the tip into the conduction band, a co-tunneling process exists in which a spin-flip exchange of a tip electron with the magnetic moment occurs while tunneling into the conduction band. They demonstrated that while the differential conductance in general exhibits a hard hybridization gap, disorder will lead to a finite quasi-particle lifetime, that renders this gap soft [see Figs.~\ref{fig:Kondo_Lattice}(a)]. Complementary to this study, Woelfle {\it et al.} \cite{Wol10} argued that it is inelastic electron-electron scattering arising from the strong Coulomb repulsion in the $f$-levels that induces a finite quasi-particle lifetime, and a subsequent softening of the hybridization gap. A similar effect was also found in DMFT studies \cite{Ben11} which have shown that incoherent processes become more important with increasing temperature, leading to a smearing out of the hybridization gap and the QPI spectrum. A similar effect also arises from the interaction of conduction or $f$-electrons with phonons \cite{Rac10}, which can lead to a complete destruction of the heavy Fermi liquid state. We note in this regard that the existence of a hard [Fig.~\ref{fig:Kondo_Lattice}(c)] or soft [Fig.~\ref{fig:Kondo_Lattice}(d)] gap in $dI/dV$ \cite{Fig10} -- omitting for a moment finite quasi-particle lifetime effects which are expected to be small at temperatures well below the coherence temperature -- depends on the existence or lack of an indirect hybridization gap in the heavy Fermi liquid bandstructure.

\subsection{Quasi-Particle Interference in Heavy Fermion Materials}
\label{sec:AndMod2}

Quasi-particle interference spectroscopy has been greatly successful in providing insight into the electronic structure of simple metals \cite{Crom93,Has93} as well as unconventional superconductors \cite{Hoff02,Han10}. Its basic idea is that defects or impurities elastically backscatter a particle with momentum ${\bf k}$ into a state with momentum ${-\bf k}$ (for electrons near the Fermi surface, this process is known as $2k_F$-scattering). Since the momentum depends on the energy of the particle -- for a free electron gas, one has $|{\bf k}(E)| = \sqrt{2m (E+\mu)}$, where $m$ is the mass of the particle, and $\mu$ the chemical potential --  this backscattering process gives rise to spatial oscillations in the energy-resolved local density of states with wave-length $\lambda = 2 \pi /(2 |{\bf k}(E)|)$. Hence by Fourier transforming the spatially resolved differential conductance $dI({\bf r},V)/dV$ -- which in a system with a single electronic band is proportional to the local density of states -- into momentum space, one gains insight into the variation of $|{\bf k}|$ with $E$, and hence the electronic dispersion of the system.

\begin{figure}[h]
\includegraphics[height=7.0cm]{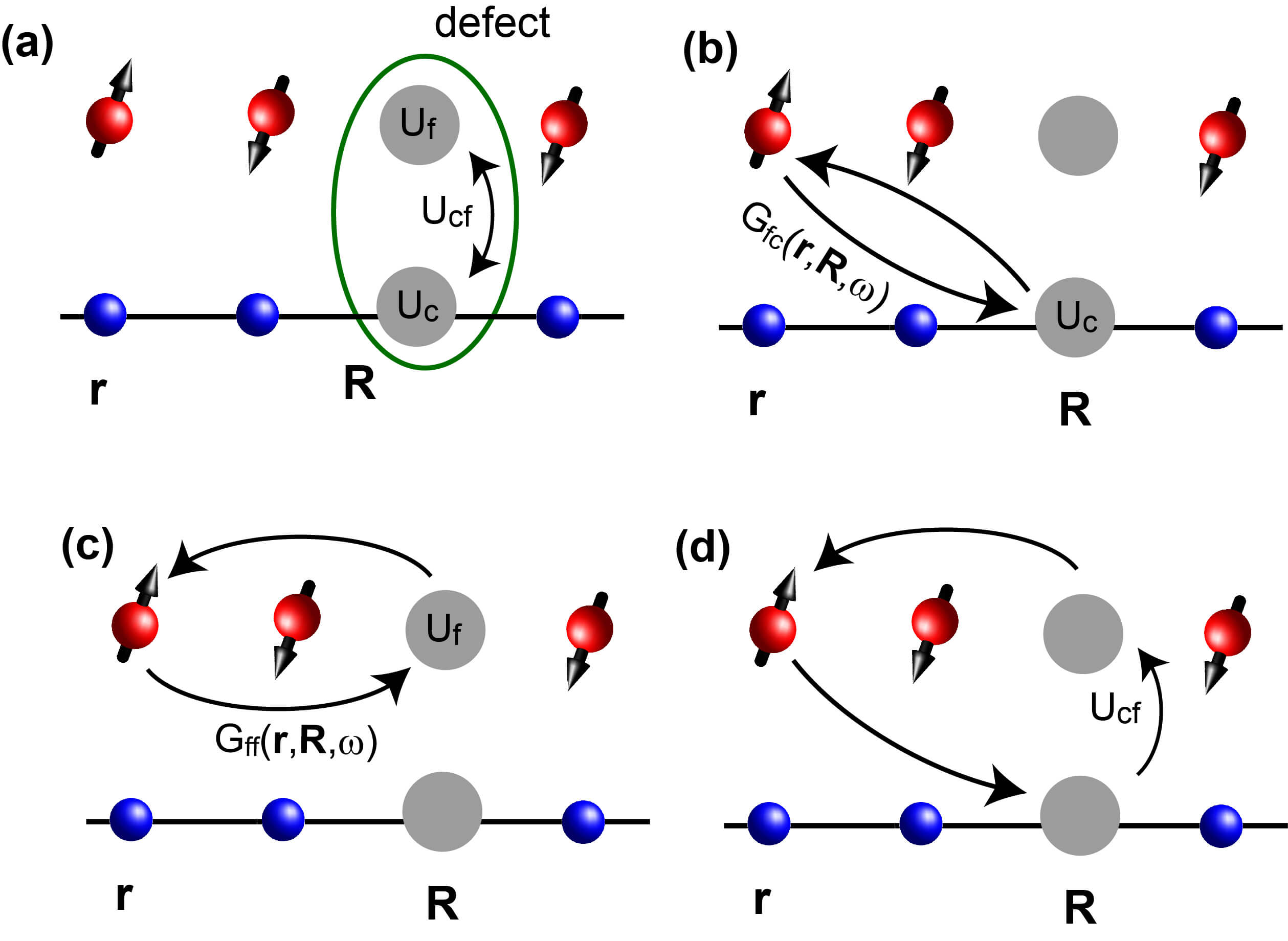}
\caption{(a) Schematic picture of a defect in a heavy fermion material, leading to intraband and interband scattering. First order scattering processes involving (b) $U_c$, (c) $U_f$, and (d) $U_{cf}$.} \label{fig:Scattering_Interference}
\end{figure}

The question naturally arises of whether QPI spectroscopy can also provide insight into the more complex electronic structure of heavy fermion materials which possess at least two different electronic bands. To examine this question, one considers the elastic scattering by static impurities described by the Hamiltonian
\begin{align}
{\cal H}_{scatt} & = \sum_{{\bf r}, \sigma} U_c c^\dagger_{{\bf r}, \sigma} c_{{\bf r}, \sigma} + U^{(0)}_f f^\dagger_{{\bf r}, \sigma} b_{\bf r} f_{{\bf r}, \sigma} b^\dagger _{\bf r} + U^{(0)}_{cf} \left( f^\dagger_{{\bf r}, \sigma} b_{\bf r} c_{{\bf r}, \sigma} + H.c. \right) \ , \label{eq:scatt}
\end{align}
where the sum runs over all impurity locations. The first two terms describe the intra-band scattering within the $c$- and $f$-electron bands, while the last term represents inter-band scattering, as schematically shown in Fig.~\ref{fig:Scattering_Interference}(a).
Within the saddle-point approximation, the effective scattering potentials are given by  $U_f=U_f^{(0)} r_0^2$ and $U_{cf}=U_{cf}^{(0)} r_0$.
Using the Born approximation, we consider only the changes in the differential conductance, $\delta \left( dI({\bf r},E=eV)/dV \right)$ to lowest (first) order in the scattering potentials. Fourier transform of $\delta \left( dI({\bf r},E=eV)/dV \right)$ into momentum space, then yields the quasi-particle interference spectrum
\begin{align}
g({\bf q},\omega) &\equiv \delta \left( \frac{dI({\bf q},\omega)}{dV} \right) = \frac{
\pi e^2}{\hbar} N_t \sum_{\sigma=\uparrow,\downarrow} \sum_{i,j=1}^2 \left[{\hat t} {\hat N}_\sigma ({\bf
q},\omega)
{\hat t} \right]_{ij} \ , \label{eq:QPI}
\end{align}
where
\begin{align}
{\hat N}_\sigma ({\bf q},\omega)&=-\frac{1}{\pi} \ {\rm Im} \int \frac{d^2
k}{(2 \pi)^2} {\hat G}_\sigma ({\bf k}, \omega) {\hat U} {\hat G}_\sigma ({\bf
k+q}, \omega)  , \quad {\rm with} \ \
{\hat U} =
\begin{pmatrix}
U_c& U_{cf} \\
U_{fc} & U_f
\end{pmatrix} \ .
\end{align}
Note that it is the quantum interference between the scattering processes associated with each of the scattering potentials, $U_c, U_f, U_{cf},$ and $U_{fc}$ (as schematically shown in Figs.\ref{fig:Scattering_Interference}(b)-(d) for the scattering of $f$-electrons off a defect) that determines the form and spectral weight distribution in the resulting QPI spectrum.

\section{Quantum Interference and Differential Conductance for a Single Kondo Impurity}
\label{sec:QI_SingleKondo}

\begin{figure}[h]
\includegraphics[width=8.5cm]{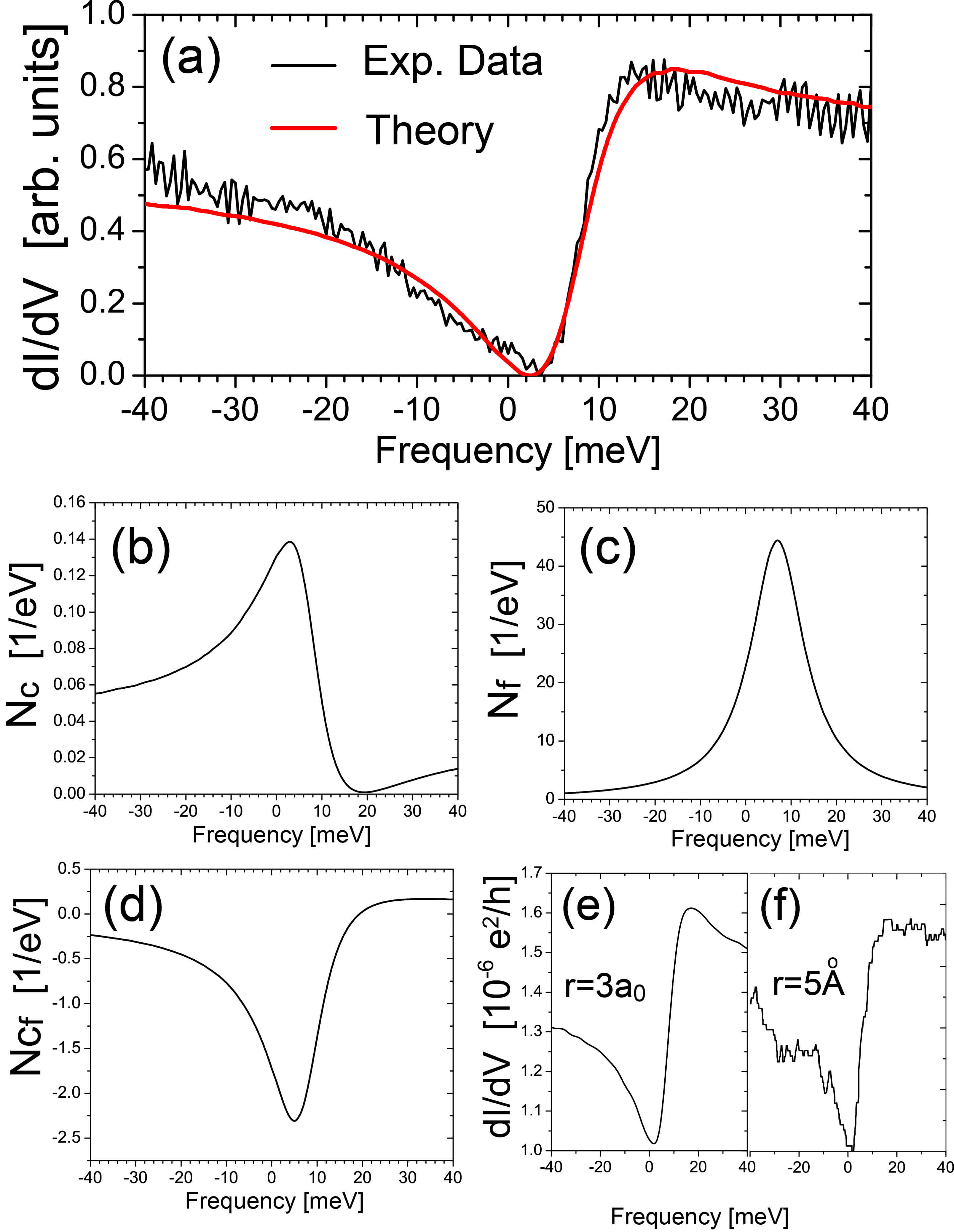}
\caption{(a) Experimental $dI/dV$ curve \cite{Mad01} at the site of a Co atom located on a Au(111) surface
together with a theoretical fit using Eq.(\ref{eq:IV}) \cite{Fig10}. A constant background was
subtracted from the experimental data. (b) Conduction electron LDOS
$N_c(\omega)$, (c) $f$-electron LDOS $N_f(\omega)$, and (d)
$N_{cf}(\omega)$ at the site of the Co atom. (e) Theoretical $dI/dV$ at a distance of $r=3
a_0$ from the Co atom for $t_f=0$. (f) Experimental $dI/dV$ curve of Ref.~\cite{Mad98}
at $r=5 \AA$ from the Co atom.} \label{fig:ExpFit}
\end{figure}

The Kondo screening of an isolated magnetic impurity is a local process that involves conduction electrons up to a distance of the size of the Kondo screening cloud from the defect \cite{Bor07,Aff01}. As scanning tunneling spectroscopy is a local probe, it is ideally suited to provide detailed insight into the complex electronic structure around the magnetic impurity, which reflects the hybridization between the conduction electron states and the state containing the magnetic moment. Madhavan {\it et al.}~\cite{Mad01} therefore investigated the form of the differential conductance in the vicinity of a Co atom located on a metallic
Au(111) surface (we note that though the magnetic moment of Co is located in a $d$-orbital, we will keep the notation of Eq.(\ref{eq:slave}) and refer to the pseudo-fermion states representing the magnetic moment as $f$-electron states). Fig.~\ref{fig:ExpFit}(a) shows the experimental $dI/dV$ data \cite{Mad01}, taken when the STS tip is positioned above a magnetic Co atom. As $T<T_K$, the $dI/dV$ data exhibit a characteristic hump-dip-peak structure which is a direct signature of the hybridization between
the conduction band and the magnetic $f$-electron state of the Co impurity -- and hence of the screening of the local moment --
and is commonly referred to as the {\it Kondo resonance}.
Overlain on the experimental results is a theoretical fit obtained by Figgins {\it et al.} \cite{Fig10} from
Eq.(\ref{eq:IV}). This fit assumes that the screening conduction band is given by the
Au(111) surface states \cite{Schout09}, and uses $N=4$ as required for the description of the $S=3/2$-spin of Co.
With this input, the theoretically computed differential conductance is entirely determined by the strength of the Kondo coupling,
$J$, and the ratio of the tunneling amplitudes $t_f/t_c$. While the former controls the width of the Kondo resonance, the latter governs its asymmetry. Note that even though the STS tip positioned above the Co atom, the extracted value of $t_f/t_c=0.066$ is small, likely due to the strong Coulomb repulsion in the $f$-level suppressing the tunneling process of an electron from the STS tip into this state.

Fig.~\ref{fig:ExpFit}(b) shows the LDOS of the conduction electrons, corresponding to $dI/dV$ in the limit $t_f/t_c = 0$. Its asymmetry is inconsistent with, and indeed opposite to the experimentally observed one shown in Fig.~\ref{fig:ExpFit}(a). Similarly, the LDOS of the $f$-electron state [see Fig.~\ref{fig:ExpFit}(c)], corresponding to $dI/dV$ in the limit $t_f/t_c \rightarrow \infty$, exhibits a single peak, and is therefore also qualitatively different from the $dI/dV$ lineshape observed experimentally. Figgins {\it et al.} \cite{Fig10} therefore concluded that the inclusion of both tunneling paths, and in particular that of the interference term $N_{cf}(\omega)$ shown in Fig.~\ref{fig:ExpFit}(d), is crucial in explaining the experimentally measured $dI/dV$ curves. Moreover, as the STS tip is moved away from the Co atom, direct tunneling into
the magnetic $f$-electron state becomes suppressed and hence $t_f \rightarrow 0$ \cite{Kno02,Ali05}. The theoretical $dI/dV$ lineshape at a distance of
$r=3 a_0$ from the Co atom [see Fig.~\ref{fig:ExpFit}(e)] obtained with $t_f=0$, shows the same asymmetry as the one at the site of the Co
atom, and qualitatively agrees with the experimental $dI/dV$ curve at $r=5 \AA$ \cite{Mad98} shown in Fig.~\ref{fig:ExpFit}(f).

\begin{figure}[h]
\includegraphics[height=3.5cm]{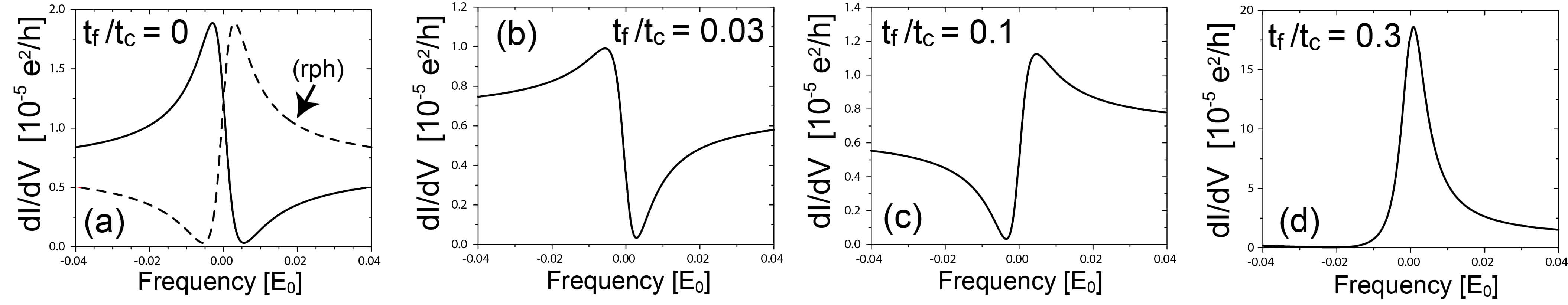}
\caption{(a) - (d) Evolution of $dI/dV$ at the location ${\bf R}$ of a magnetic atom with increasing $t_f/t_c$ \cite{Fig10}. Dashed line in (a) represents $dI/dV$
for a conduction band with a reversed particle-hole asymmetry.} \label{fig:Fig2PRL}
\end{figure}
The microscopic origin of the asymmetry in the $dI/dV$ lineshape does not only lie in the
existence of two tunneling paths, but also in the particle-hole asymmetry of the screening
conduction band. To demonstrate this, Figgins {\it et al.} \cite{Fig10} considered the case of a single magnetic impurity with a spin-$1/2$
moment, corresponding to $N=2$, and a conduction band whose Fermi wavelength $\lambda_F = 10 a_0$ is representative
of the Au(111) and Cu(111) surfaces states \cite{Schout09}. For $t_f=0$ [solid line in Fig.~\ref{fig:Fig2PRL}(a)], the $dI/dV$ lineshape exhibits
a Kondo resonance whose asymmetry is a direct consequence of the particle-hole
asymmetry of the conduction band. Indeed, reversing the latter via
$\mu \rightarrow -\mu$, also leads to a reversal of the asymmetry in
$dI/dV$ [see dashed line in Fig.~\ref{fig:Fig2PRL}(a)], thus demonstrating the effect of the conduction band's particle-hole asymmetry on the $dI/dV$ lineshape. With increasing $t_f/t_c$, the $dI/dV$ lineshape undergoes a characteristic evolution, in which its asymmetry is
first reversed [Figs.~\ref{fig:Fig2PRL}(b) - \ref{fig:Fig2PRL}(c)], and subsequently, its characteristic peak-dip-hump structure is replaced by a single (asymmetric) peak [Fig.~\ref{fig:Fig2PRL}(d)]. The latter is a clear indication that as $t_f/t_c$ becomes sufficiently large, the main contribution to $dI/dV$ arises from the magnetic $f$-level.

\section{Differential Conductance and Quasi-Particle Interference in Heavy Fermion Materials}
\label{sec:HF}

\subsection{The Hidden Order Phase of URu$_2$Si$_2$}
\label{sec:URuSi}

One of the most puzzling heavy fermion materials is URu$_2$Si$_2$ which possesses a coherence temperature of
$T_{coh} \approx 55$K \cite{Pal85,Map86} and
undergoes a second order phase transition at $T_0 =17.5$K
\cite{Pal85,Map86,Scho87,Bonn88,Dor01,Rod97,Esc94,Den01} into a state whose microscopic nature is still unknown, and which is therefore called the {\it hidden
order phase}.
\begin{figure}[h]
\includegraphics[height=3.5cm]{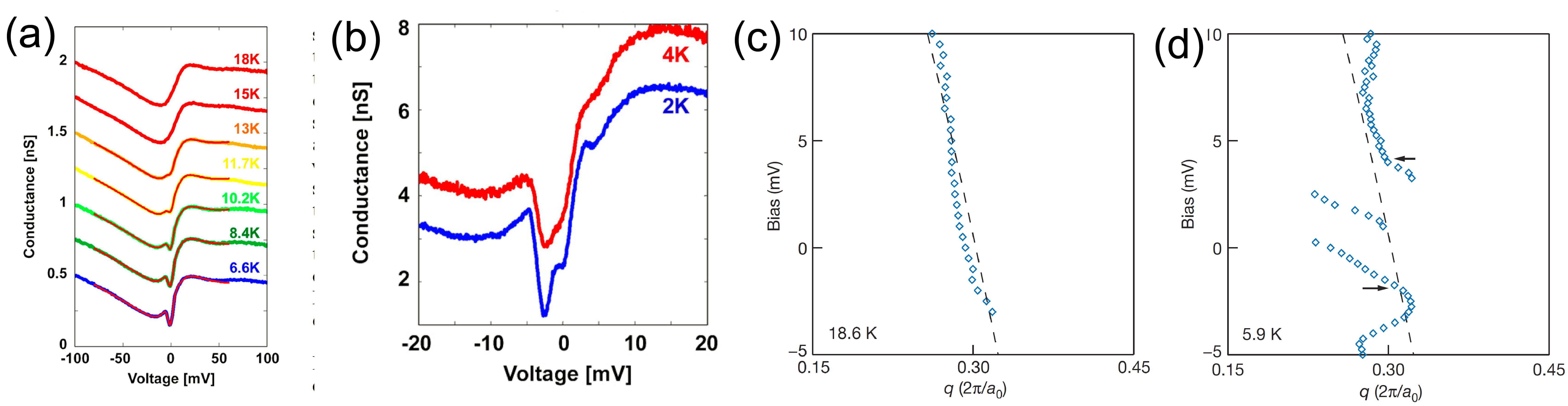}
\caption{(a),(b) Temperature evolution of the differential conductance, $dI/dV$ through the hidden order transition at $T_0$ \cite{Ayn10} in pristine URu$_2$Si$_2$. Evolution of the QPI spectrum from (c) above $T_0$ to (d) below $T_0$ in Th-doped URu$_2$Si$_2$ \cite{Sch09}.} \label{fig:URuSi_exp}
\end{figure}
While the debate on the nature of this state is still ongoing \cite{Map86,Cox87,Cha02,Var06,Bal09,Elg09,Pep11,Hau09,Opp10,Dubi11,Cha13,Myd14,Oka11,Kung15},
new insight into this question has been provided by a series of scanning tunneling
spectroscopy experiments \cite{Sch09,Ayn10} [see Fig.~\ref{fig:URuSi_exp}]. These experiments have shown that $dI/dV$ exhibits the opening of a soft gap below $T_0$ [see Figs.~\ref{fig:URuSi_exp}(a) and (b)] \cite{Sch09,Ayn10}, and that the QPI dispersion -- corresponding to that ${\bf q}$ at which $|g({\bf q},E)|$ exhibits a maximum for fixed $E$ -- significantly evolves through $T_0$ [see Figs.~\ref{fig:URuSi_exp}(c) and (d)]. In particular, the QPI dispersion exhibits a form at $T \ll T_0$, which was suggested to be, at least qualitatively, consistent with that in the heavy Fermi liquid phase \cite{Sch09} of a screened Kondo (or Anderson) lattice.

\begin{figure}[h]
\includegraphics[width=8.cm]{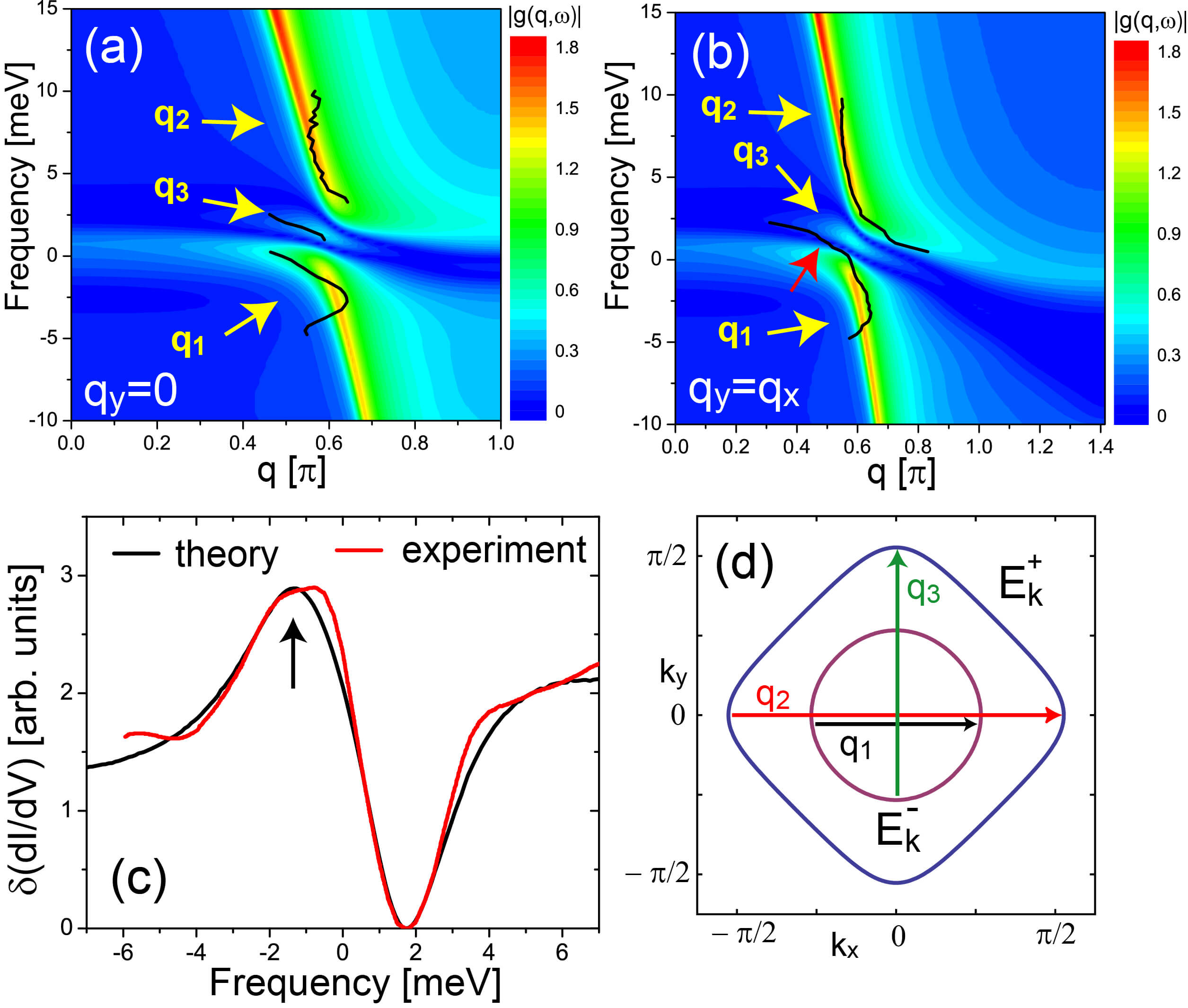}
\caption{Contour plot of the theoretical $|g({\bf q},\omega)|$ \cite{Yuan12} along
(a) $q_y=0$ and (b) $q_y=q_x$, together with the experimental QPI dispersions (black lines) \cite{Sch09}. (c) Experimental \cite{Sch09} and
theoretical \cite{Yuan12} $\delta(dI/dV)$ below $T_0$. (d) Fermi surfaces
of $E_{\bf k}^{\pm}$ obtained from the theoretical fit \cite{Yuan12} of the experimental QPI data  \cite{Sch09}.} \label{fig:Fig1PRB}
\end{figure}

Yuan {\it et al.} \cite{Yuan12} proposed a theoretical model to analyse these experimental findings and argued that they \cite{Sch09,Ayn10} reflect the
emergence of a coherent Anderson lattice, and hence a heavy Fermi liquid state, below the hidden order transition. Their first evidence for this conclusion comes from the theoretical fits [see Figs.~\ref{fig:Fig1PRB}(a) and (b)] of the experimental QPI dispersions (black lines) measured by Schmidt {\it et
al.}~\cite{Sch09} on a U-terminated surface of a 1\% Th-doped URu$_2$Si$_2$ sample \cite{Sch09}. In this sample, it is the Th atoms that scatter the conduction electrons, and induce the spatial oscillations in $dI/dV$ that are necessary to obtain a QPI spectrum. The theoretically computed contour plots of $|g({\bf q},\omega)|$ -- obtained with $U_f/U_c \approx 0.6$ and $U_{cf}=0$ from Eq.(\ref{eq:QPI}) -- in Figs.~\ref{fig:Fig1PRB}(a) and (b) reflect the existence of two hybridized bands, characteristic of the heavy Fermi liquid state. The good agreement between the maxima in the theoretical QPI contour plots and the experimental QPI dispersions allowed Yuan {\it et al.} \cite{Yuan12}  to extract the momentum structure of the unhybridized bands, $\varepsilon^{c,f}_{\bf k}$, of the hybridization, $s$, and of the hybridized bands, $E_{\bf k}^{\pm}$. This, in turn, enabled them to compute the change in $dI/dV$ below $T_0$, i.e., $\delta(dI/dV)=dI/dV(T<T_0)-dI/dV(T=T_0)$ \cite{Yuan12}, which is shown in Fig.~\ref{fig:Fig1PRB}(c) together with the experimental result \cite{Sch09}.
Yuan {\it et al.} argued that the good quantitative agreement between
the theoretical and experimental $dI/dV$ lineshapes and QPI dispersions, and the consistency between these two sets of data,
strongly suggests that the STS data reflect the existence of a heavy Fermi liquid state in the form of a coherent Anderson lattice of screened magnetic moments below $T_0$, confirming the proposal made by Schmidt et al.\cite{Sch09}.

Yuan {\it et al.} further argued that the form of the QPI spectrum is determined by scattering of electrons both within and between the $E_{\bf k}^{\pm}$-bands with intraband
scattering [see Fig.~\ref{fig:Fig1PRB}(d)] giving rise to the ${\bf q}_1$ and ${\bf q}_2$ branches in $|q({\bf q},\omega)|$ shown in
Figs.~\ref{fig:Fig1PRB}(a) and (b). The overlap of the energy dispersions, $E_{\bf k}^{\pm}$, of the two hybridized bands in the energy interval $ -1 \mbox{ meV} \lesssim \omega \lesssim 1.5$ meV, allows for interband scattering with wave-vector ${\bf q}_3$, and a
corresponding ${\bf q}_3$ branch in $|q({\bf q},\omega)|$ [see
Figs.~\ref{fig:Fig1PRB}(a) and (b)]. The ${\bf q}_3$ branch was observed experimentally along $q_y=0$, thus confirming the theoretical prediction,
but not along $q_y=q_x$. This ``missing" branch is likely due to the smaller separation between the branches along
this direction rendering the experimental resolution of the ${\bf q}_1$
and ${\bf q}_3$ branches difficult. The agreement between the theoretical and experimental QPI dispersions also provides further insight into the form of
$dI/dV$ in that it identifies the  peak in $dI/dV$ at $\omega = -2$ meV [see arrow in
Fig.~\ref{fig:Fig1PRB}(c)] as arising from the van Hove singularity of the $f$-electron band.

\begin{figure}[h]
\includegraphics[width=8.cm]{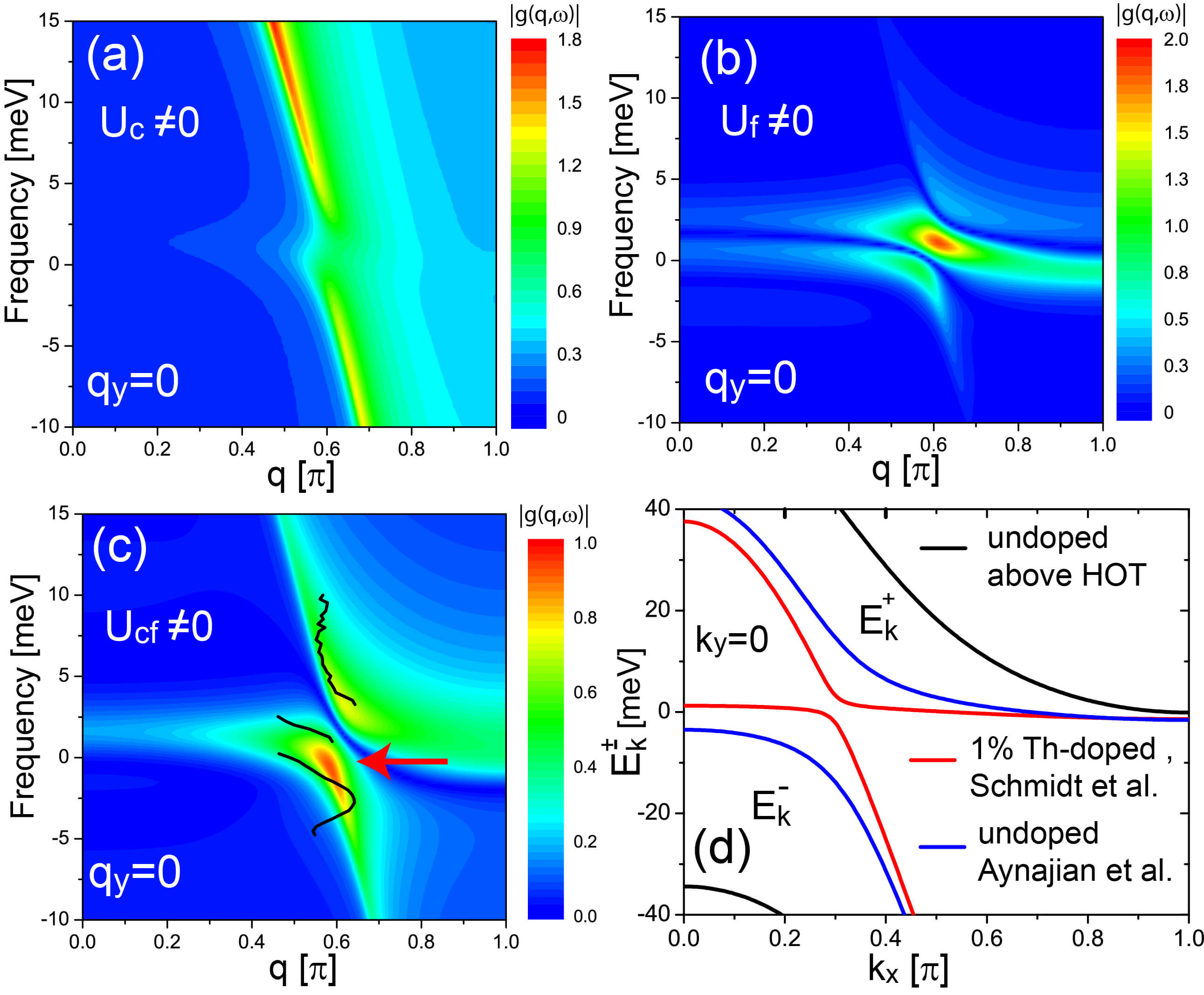}
\caption{ $|g({\bf q},\omega)|$ along $q_y=0$ for (a) $U_f,U_{cf}=0$, $U_c \not = 0$, (b) $U_c,U_{cf}=0$, $U_f \not = 0$, and (c)
$U_f,U_c=0$, $U_{cf} \not = 0$, together with the QPI dispersions of
Ref.~\cite{Sch09} (black lines). (d) $E_{\bf
k}^\pm$ extracted from the theoretical fits \cite{Yuan12}.  } \label{fig:Fig2a}
\end{figure}

The experimental QPI spectra \cite{Sch09} also provide insight
into the microscopic mechanism underlying the electronic scattering by Th atoms, as the
spectral weight associated with the QPI spectrum $|q({\bf q},\omega)|$ sensitively depends on the relative strength of the scattering potentials, and hence the quantum interference between the scattering channels. To demonstrate this, Yuan {\it et al.} \cite{Yuan12} contrasted the QPI spectra obtained when only one of the three scattering potentials, $U_c, U_f$ and $U_{cf}$ is non-zero [see Figs.~\ref{fig:Fig2a}(a) - (c)].
For intraband scattering with $U_c \not = 0$ [Fig.~\ref{fig:Fig2a}(a)] and $U_f \not = 0$ [Fig.~\ref{fig:Fig2a}(b)] the dominant contribution to the QPI spectrum arises from
scattering between those states where the coherence factors of the $c$-electrons and $f$-electrons, respectively, are large. However, as the spectral weight in $|q({\bf q},\omega)|$ in both cases is inconsistent with the experimentally observed QPI weight and dispersion, the latter can only be explained (as shown in Fig.~\ref{fig:Fig1PRB}) by considering intraband scattering in both the $c$- and $f$-electron bands with relative scattering strength $U_f/U_c \approx 0.6$. Moreover, for interband scattering between the $c$ and $f$-bands, $U_{cf} \not = 0$ [Fig.~\ref{fig:Fig2a}(c)], the QPI spectrum significantly deviates
from the experimental QPI dispersion. In particular, interband scattering leads to only two branches in the QPI spectrum, in contrast to the three branches observed experimentally. In addition, the largest spectral weight in the QPI spectrum occurs where the experimental QPI intensity is close to a minimum [see red arrow in Fig.~\ref{fig:Fig2a}(c)]. These inconsistencies thus strongly suggest that the interband scattering by Th-atoms is negligible.

\begin{figure}[h]
\includegraphics[height=7.0cm]{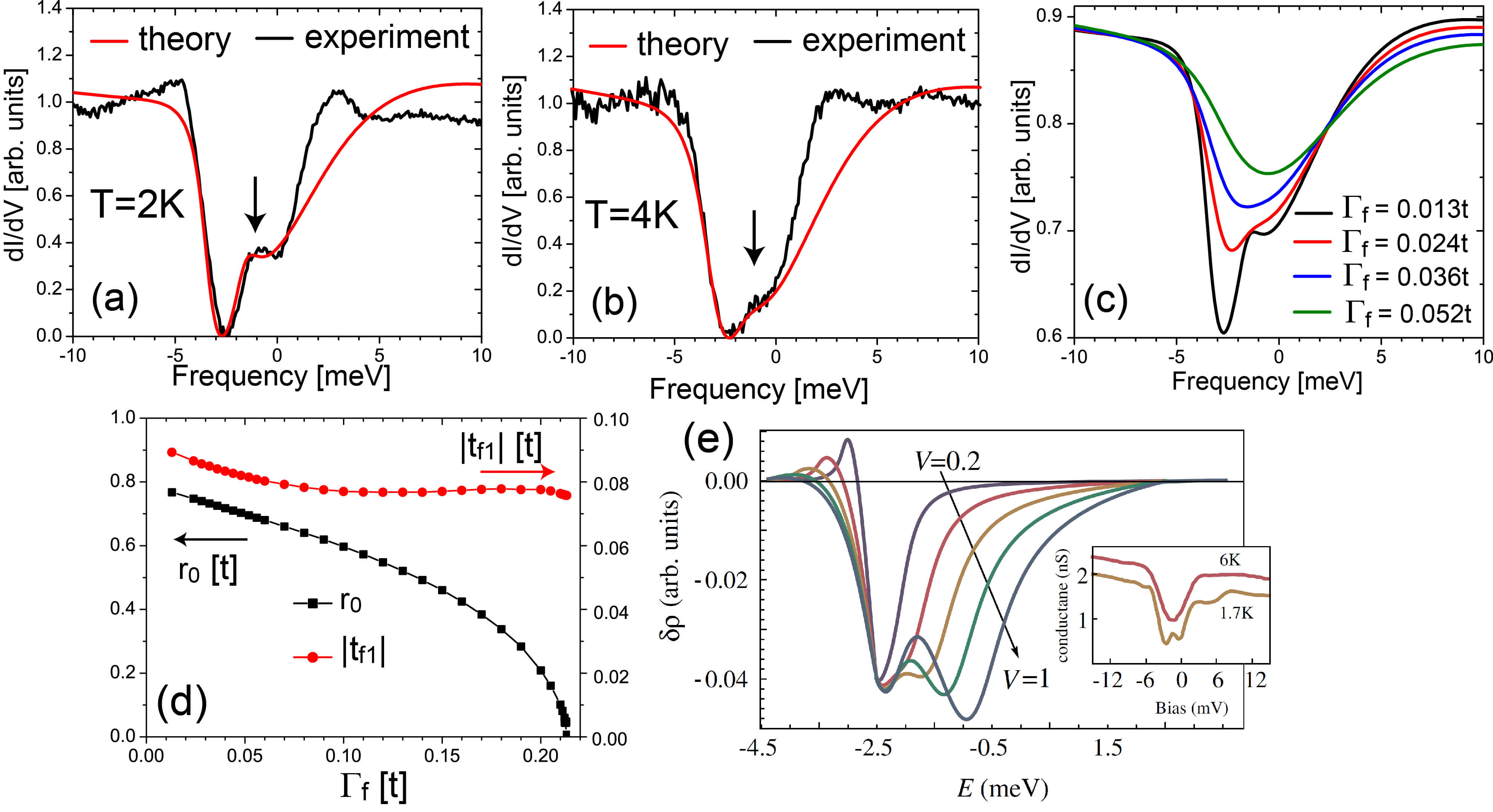}
\caption{Theoretical fits \cite{Yuan12} to the $dI/dV$ data of
Ref.~\cite{Ayn10} at (a) $T=2$K and (b) $T=4$K. (c) Evolution of
$dI/dV$ with $\Gamma_f$. (d) Dependence of $r_0$ and $|t_{f1}|$ on $\Gamma_f$. (e) Evolution of $dI/dV$ with increasing order parameter of the hybridization wave in URu$_2$Si$_2$ \cite{Dubi11}.} \label{fig:Fig2}
\end{figure}

Further evidence for the existence of a heavy Fermi liquid state below $T_0$ comes from the differential conductance measured by Aynajian {\it et al.} \cite{Ayn10}  on a U-terminated surface of pure URu$_2$Si$_2$ at $T=2$K and $4$K, respectively [see Figs.~\ref{fig:Fig2}(a) and (b)]. Starting from their analysis of the QPI spectra by Schmidt {\it et al.} \cite{Sch09}, Yuan {\it et al.}\cite{Yuan12} argued that the theoretical fits of the experimental $dI/dV$ lineshapes reproduce all of
the experiment's salient features: the asymmetry and
magnitude of the gap in $dI/dV$ as well as the peak at $\omega
\approx -0.8$ meV [see arrows in Figs.~\ref{fig:Fig2}(a) and (b)]
which arises from the van Hove singularity of the $f$-electron band. Similar features were also observed by Schmidt {\it et al.} \cite{Sch09}.
As these features are characteristic signatures of the hybridized band structure in the heavy Fermi liquid state, the experimental $dI/dV$ lineshapes provide further evidence for its existence.

The temperature evolution of $dI/dV$ observed by Aynajian {\it et al.} \cite{Ayn10} [see Figs.~\ref{fig:URuSi_exp}(a) and (b)] also allows one to gain insight into the
microscopic mechanism that drives the emergence of the heavy Fermi liquid state below $T_0$. To this end, Yuan {\it et al.} showed that the observed changes in
$dI/dV$ between $T=2$K [Fig.~\ref{fig:Fig2}(a)] and $T=4$K [Fig.~\ref{fig:Fig2}(b)] can be solely attributed to an increasing decoherence (as described by the decoherence rate $\Gamma_f$) of the $f$-electron states. Increasing $\Gamma_f$ even further yields the evolution of
$dI/dV$ shown in Fig.~\ref{fig:Fig2}(c) which possesses the same
characteristic signatures as those observed by Aynajian {\it et al.} \cite{Ayn10} with increasing temperature [see Fig.~\ref{fig:URuSi_exp}(a)]: the gap in
$dI/dV$ is filled in, its magnitude remains approximately constant until one approaches the hidden order transition, and the center of the gap shifts to larger
energies (a similar temperature dependence was also found by Schmidt {\it et al.} \cite{Sch09}). It is instructive to consider the evolution of $r_0$ and $t_{f1}$ [see Fig.~\ref{fig:Fig2}(d)] with increasing $\Gamma_f$, as obtained from the self-consistent solution
of Eqs.(\ref{eq:Andsc1}) - (\ref{eq:Andsc1}). While $t_{f1}$ varies only weakly with increasing $\Gamma_f$, $r_0$, and hence the effective hybridization $s$, is strongly suppressed and eventually vanishes at $\Gamma_{f}=\Gamma^c_{f}$. These results, taken together, strongly suggest that the experimentally observed formation of a heavy Fermi liquid state below $T_0$ is driven by a significant reduction in $\Gamma_f$ at $T_0$ from $\Gamma_f > \Gamma^c_{f}$ above $T_0$ to $\Gamma_f < \Gamma^c_{f}$ below $T_0$. Chatterjee {\it et al.} \cite{Chatt13} recently arrived at a similar conclusion based on the results of photoemission experiments. These conclusions demonstrate the importance of inelastic scattering process (giving rise to $\Gamma_f$) in the destruction of the coherent Anderson lattice \cite{Ben11,Rac10}. An alternative explanation was proposed by Dubi and Balatsky \cite{Dubi11}. They argued that a hybridization wave emerges below $T_0$ arising from the particle-hole pairing of an $f$-hole with momentum $Q=0.3\pi/a_0$ and a $c$-electron with momentum $-Q$. With increasing strength of the order parameter, $V$, the resulting $dI/dV$ [Fig.\ref{fig:Fig2}(e)] develops a gap and for sufficiently large $V$, also exhibits a peak inside the gap, consistent with the experimental findings \cite{Sch09}.

While Yuan {\it et al.} \cite{Yuan12} concluded that the experimental STS data do not exhibit a direct signature
of a hidden order parameter below $T_0$, but rather reflect the existence of a coherent heavy Fermi liquid, they argued that the deduced strong reduction in $\Gamma_f$ at $T_0$ might be a direct signature of this order parameter. In particular, if $\Gamma_f$ arises due to a coupling of the $f$-electrons to a fluctuating mode associated with the hidden order parameter, then the condensation of the order parameter at $T_0$ would significantly reduce the electron-mode coupling, thus reducing the decoherence of the $f$-electrons, as reflected in a suppression of $\Gamma_f$. A recent alternative explanation, ascribing the hidden order phase to the emergence of a {\it hastatic order} \cite{Cha13}, has proposed that the hidden order parameter might be ``hidden" in the detailed momentum dependence of the hybridization  between the $f$- and $c$-electrons. A test of this scenario will require a detailed comparison between the $dI/dV$ lineshapes and QPI spectra following from this proposal with the experimental STS results.

Finally, a comparison of the bandstructure, $E_{\bf k}^\pm$, [Fig.~\ref{fig:Fig2a}(d)] extracted from the STS experiments on the  1\% Th-doped \cite{Sch09} and pristine URu$_2$Si$_2$ \cite{Ayn10} samples, has shown that Th-doping decreases the hybridization, and hence the hybridization gap, and increases $\Gamma_{f,c}$, and hence the decoherence of the quasi-particles. Moreover, knowledge of the bandstructure allows one to explore the origin of the magnetic interaction, $I_{{\bf r,r'}}$. While Yuan {\it et al.} showed that the extracted $I_{{\bf r,r'}}$ between nearest-neighbor sites is antiferromagnetic, the magnetic RKKY-interaction computed from the extracted bandstructure is ferromagnetic. This result suggests that the microscopic origin of the nearest-neighbor $I_{{\bf r,r'}}$ in URu$_2$Si$_2$ lies in direct exchange, and not in an RKKY interaction.

\subsection{Differential Conductance and QPI spectroscopy in CeCoIn$_5$}
\label{sec:CeCoIn5}

One of the most interesting heavy fermion materials is CeCoIn$_5$, a member of the so-called ``115"-family, whose phase diagram [see Fig.~\ref{fig:PD}(b)] shares many of the
fascinating features that are also found in the phase diagram of the cuprate \cite{Kei15} and iron-based superconductors \cite{Bas11}, such as unconventional
superconductivity in proximity to antiferromagnetism. CeCoIn$_5$ \cite{Pet01} exhibits the largest $T_c=2.3$K in this family of heavy fermion materials, and has long been considered the ``hydrogen atom" of heavy fermion superconductivity \cite{Ste79,Miy86,Beal86,Sca86,Sca87,Lav87,Col89}. While much experimental \cite{Iza01,Mos01,Koh01,Cur12,Sto08,Park08,Hu12,Koi13,Tru13,Shu14,Kim15} and theoretical effort \cite{Fli10,Sca12,Das13,Dav13,Liu13,Yang14,Wu15,Ert15} has focused on illuminating its unconventional properties \cite{Kim01,Pag03,Bia03,Pag06,How11}, and the microscopic mechanism underlying the emergence of superconductivity, no consensus has been reached to-date. A major obstacle in providing a quantitative or even qualitative explanation for its properties in the superconducting state has been the lack of insight into the material's complex electronic bandstructure \cite{Koi13}.

\begin{figure}[h]
\includegraphics[height=7.cm]{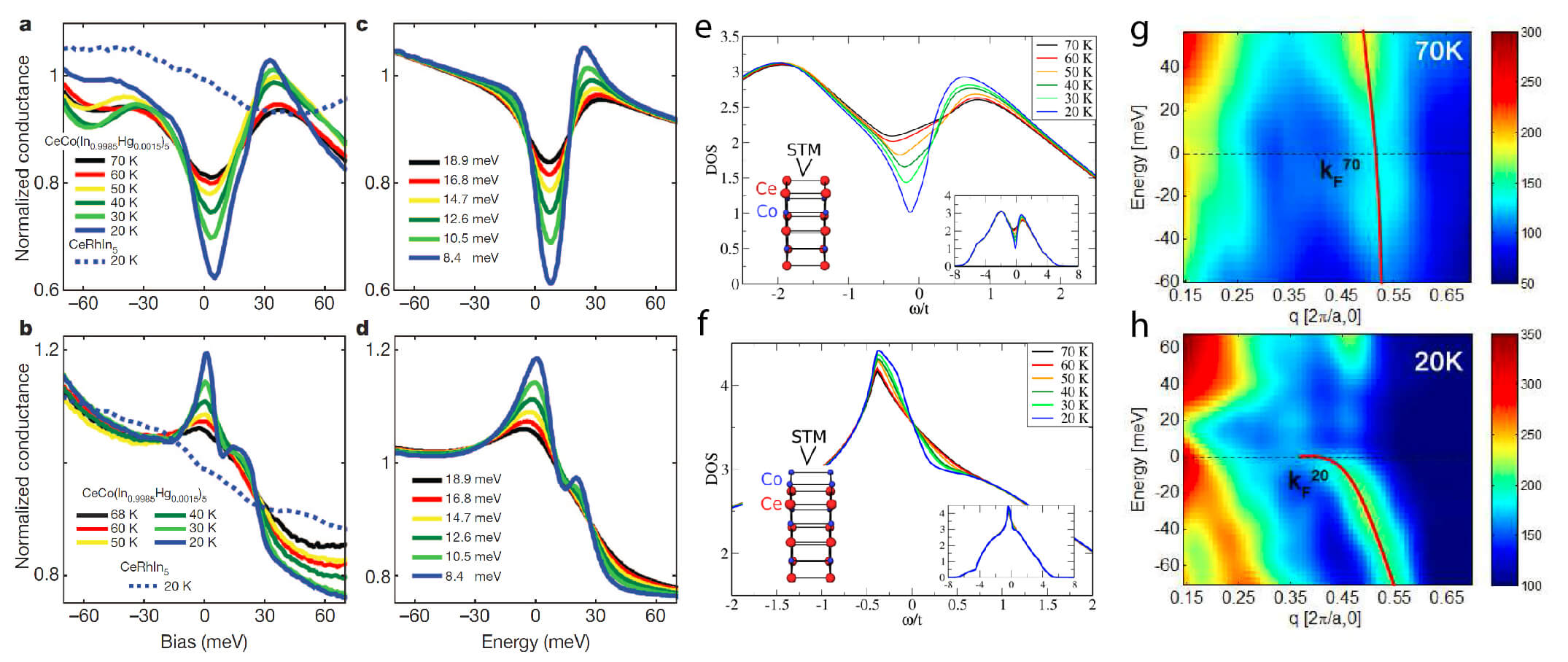}
\caption{Differential conductance \cite{Ayn12} measured on the (a) Ce-In surface layer, and (b) Co surface layer of CeCoIn$_5$. (c),(d) Theoretical fits \cite{Ayn12} of the experimental $dI/dV$ shown in (a),(b) using the formalism presented in Secs.~\ref{sec:AndMod1} and \ref{sec:AndMod2}. Theoretically computed density of states \cite{Pet14} on the (e) Ce-In surface layer, and (f) Co surface layer of CeCoIn$_5$. QPI spectrum of CeCoIn$_5$ \cite{Ayn12} at (g) $T=70$K, and (h) $T=20$K.} \label{fig:CeCoIn5}
\end{figure}
Recent STS experiments \cite{Ayn12,All13,Zhou13} have therefore focused on identifying the complex electronic bandstructure of CeCoIn$_5$ by employing quasi-particle interference spectroscopy. In particular, STS experiments by Aynajian {\it et al.}\cite{Ayn12} on CeCo(In$_{0.9985}$Hg$_{0.0015})_5$
demonstrated that below the coherence temperature, $T_{coh} \approx 45$K \cite{Park05}, but above $T_c$, the differential conductance measured on the Ce-In surface layer exhibits a typical Kondo resonance [see Fig.~\ref{fig:CeCoIn5}(a)], confirming that the material is in a heavy Fermi liquid state. A comparison of these results with those obtained on a Co termination layer show striking differences, confirming the conclusion of Sec.\ref{sec:AndMod2}, that the nature of the surface termination layer should possess a strong effect on the $dI/dV$ lineshape. In particular, while $dI/dV$ on the Ce-In surface layer, exhibits a hump-dip-peak structure, implying that electrons from the tip tunnel predominantly into the conduction band [Fig.~\ref{fig:CeCoIn5}(a)] with a correspondingly small $t_f/t_c$ [see theoretical fit in Fig.~\ref{fig:CeCoIn5}(c)], $dI/dV$ on a Co termination layer shows a strong peak, which is evidence for dominant tunneling into the heavy $f$-electron band [Fig.~\ref{fig:CeCoIn5}(b)], and hence a large value of $t_f/t_c$ [see theoretical fit in Fig.~\ref{fig:CeCoIn5}(d)]. An alternative explanation was recently put forth by Peters and Kawakami \cite{Pet14}. They proposed that it is only the conduction band orbital at the Ce site that directly couples to the Ce atom's magnetic moment, but not the conduction band orbitals at the In and Co sites. Using a DMFT approach, they qualitatively reproduced the $dI/dV$ lineshapes on the Ce-In surface [Fig.\ref{fig:CeCoIn5}(e)] and Co surface layers [Fig.\ref{fig:CeCoIn5}(f)] of CeCoIn$_5$.

Further evidence for the formation of a heavy Fermi liquid state is provided by the significant changes in the QPI spectrum that occur with decreasing temperature, as shown in Figs.~\ref{fig:CeCoIn5}(g) and (h). In particular, while at $T=70$K$>T_{coh}$, the QPI spectrum [Figs.~\ref{fig:CeCoIn5}(g)] reflects the existence of a light conduction band that crosses the Fermi energy, the onset of hybridization between the light and heavy bands at $T_{coh}$ leads to a bending of this light band, as evidence by the QPI spectrum at $T=20$K$<T_{coh}$, shown in Fig.~\ref{fig:CeCoIn5}(h).  These temperature dependent changes reflect the formation of a hybridized bandstructure characteristic for the onset of a coherent heavy Fermi liquid below $T_{coh}$. Finally, Aynajian {\it et al.}\cite{Ayn12} argued that the nearly linear temperature dependence of the width of the peak observed in $dI/dV$ on the Co surface [see Fig.~\ref{fig:CeCoIn5}(b)] is a signature of the material's proximity to a quantum critical point.

\begin{figure}[h]
\includegraphics[height=6.cm]{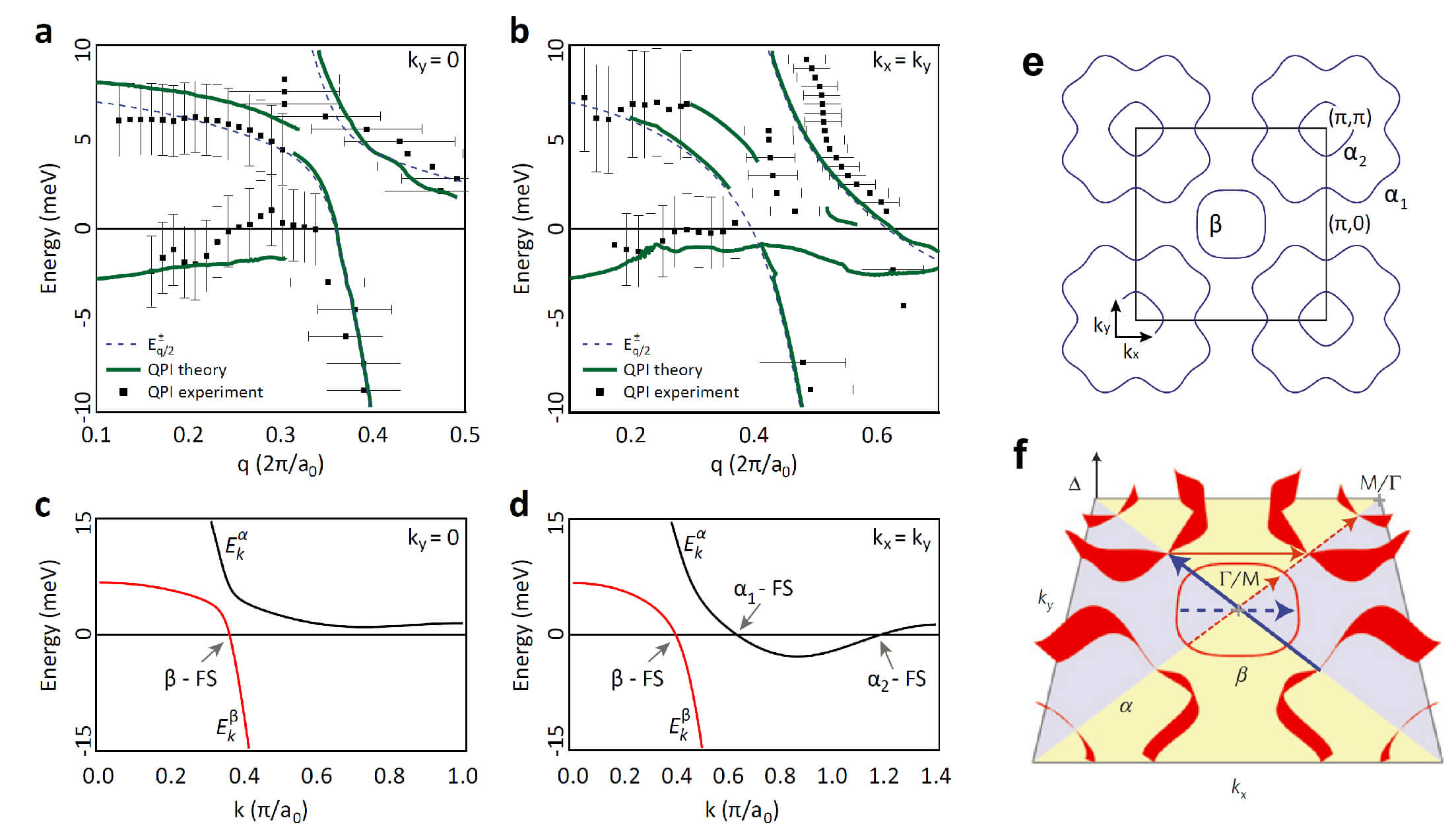}
\caption{Comparison of the experimentally measured and theoretically computed QPI dispersions along (a) $q_y=0$ and (b) $q_x=q_y$. Theoretically extracted dispersion of the hybridized bands along (c) $k_y=0$ and (d) $k_x=k_y$. The dispersion in (d) reveals as backbending of the heavy band, giving rise to the three Fermi surface sheets shown in (e). (f) Momentum structure of the extracted superconducting gaps with $d_{x^2-y^2}$-symmetry on the three Fermi surface sheets of CeCoIn$_5$ \cite{All13,Dyke14}.} \label{fig:QPICeCoIn5}
\end{figure}

More detailed insight into the complex momentum structure of the hybridized bands at a temperature $T=250$mK well below $T_{coh}$ was provided in high-resolution QPI studies by Allan {\it et al.} \cite{All13,Dyke14}. Using the theoretical formalism of Sec.~\ref{sec:AndMod2}, they obtained good agreement between the experimentally observed and theoretically computed QPI dispersions [see Figs.\ref{fig:QPICeCoIn5}(a) and (b)] that did not only reveal a momentum dependent hybridization [cf. Figs.\ref{fig:QPICeCoIn5}(c) and (d)], but also a backbending of the heavy band [Fig.\ref{fig:QPICeCoIn5}(d)], resulting in three Fermi surface sheets [see Fig.\ref{fig:QPICeCoIn5}(e)].

We briefly mention that the detailed insight into the momentum structure of the heavy bands near the Fermi surface allowed Allan {\it et al.} \cite{All13} to extend the QPI analysis into the superconducting state. Their study revealed the momentum structure of an unconventional superconducting order parameter which is consistent with a $d_{x^2-y^2}$-wave symmetry on all three Fermi surfaces, as shown in Fig.~\ref{fig:QPICeCoIn5}(f). The largest superconducting gap $\Delta_{max} \approx 0.6$ meV resides on the $\alpha_1$ Fermi surface, followed by smaller gaps on the $\alpha_2$- and $\beta$-Fermi surfaces. These results are consistent with the conclusions of Zhou {\it et al.} \cite{Zhou13} based on their QPI and $dI/dV$ data taken in the superconducting state of CeCoIn$_5$. Moreover, Dyke {\it et al.} \cite{Dyke14} were able to use this detailed insight into the complex electronic structure of CeCoIn$_5$ to extract a crucial missing component in the quest for the superconducting pairing mechanism, the pairing interaction between the magnetic $f$-moments. This interaction, together with the detailed form of the bandstructure, allowed Dyke {\it et al.} \cite{Dyke14} to solve the superconducting pairing problem, and compute a series of physical properties in the superconducting state of CeCoIn$_5$. The good agreement of their results with the experimental findings provides strong support for a superconducting pairing mechanism in CeCoIn$_5$ that is mediated by the antiferromagnetic interactions between $f$-electron moments.

\section{Hybridization Waves and Impurity States: Effect of Defects on the Local Electronic Structure of Heavy Fermion Materials}
\label{sec:Defects}

Understanding how the formation of antiferromagnetism competes
locally with the creation of a Kondo singlet, and how the magnetic
and electronic degrees of freedom are coupled, both in real and momentum space, is crucial
for identifying the microscopic mechanism underlying the complex phase diagram of heavy fermion materials.
The great success in employing defects and impurities in the high-temperature superconductors to
gain insight into their complex electronic structure \cite{Yaz99,Hud99,Bal06}, raises the question of whether a similar approach can also be
used in heavy-fermion materials to disentangle and spatially resolve
their electronic and magnetic structure.

To answer this question, Figgins {\it et al.} \cite{Fig11} investigated the spatial entanglement of electronic and magnetic degrees of freedom by exploring the effects of defects in the form of missing magnetic moments -- {\it Kondo holes} -- and non-magnetic scatterers on the local electronic structure
of heavy fermion materials.
\begin{figure}[h]
\includegraphics[height=5.0cm]{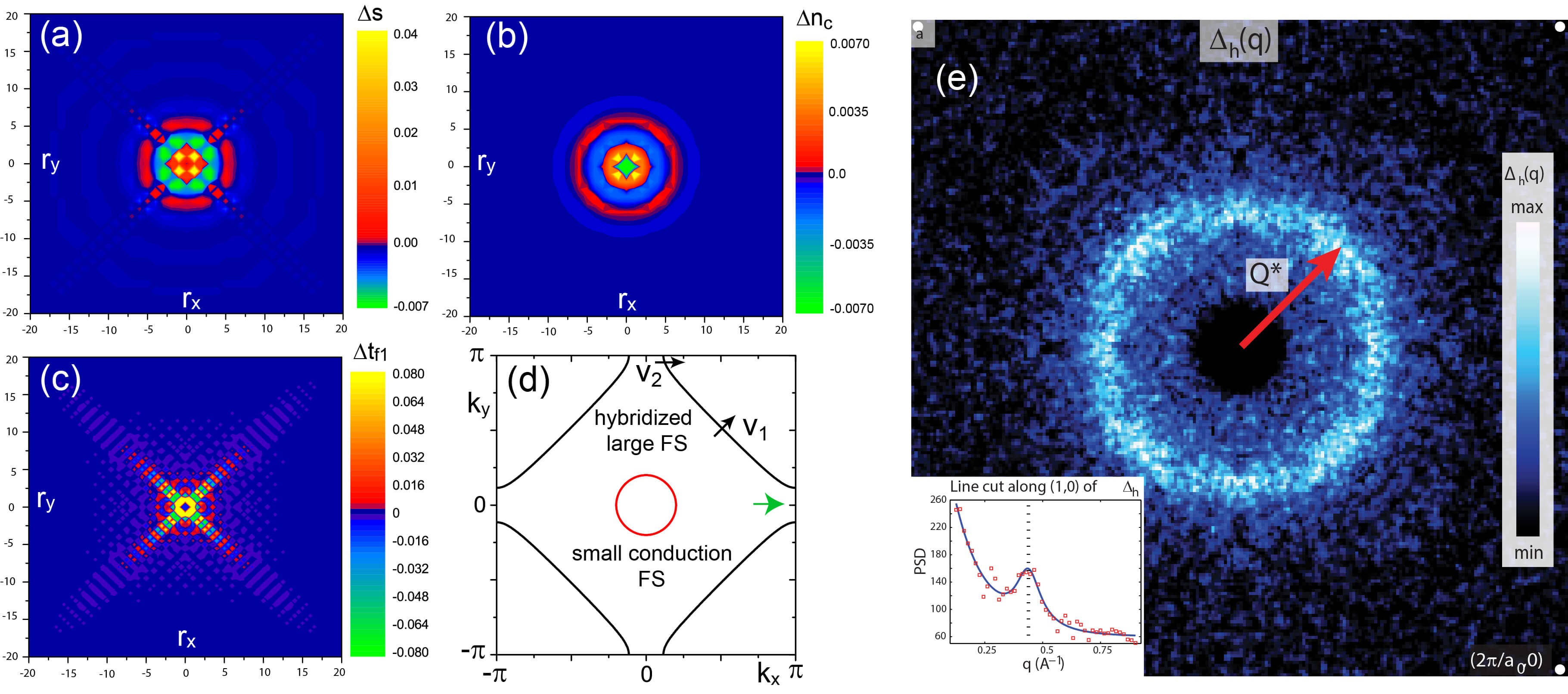}
\caption{Contour plots of (a) $\Delta s$, (b) $\Delta n_c$, and
(c) $\Delta t_{f1} ({\bf r},{\bf r}')$ [shown at $({\bf r}+{\bf
r}')/2$] in the presence of a Kondo hole located at the center. (d) Large Fermi surface (black) of the unperturbed Kondo
lattice arising from the hybridization of the $f$-electron and
conduction bands and small (red) Fermi surface of the unhybridized
conduction band. (e) Contour plot of the Fourier transformed hybridization gap map \cite{Ham11}.}
\label{fig:KH1}
\end{figure}
They showed that a Kondo hole induces significant spatial oscillations in the hybridization [see Fig.~\ref{fig:KH1}(a)], the charge density [see Fig.~\ref{fig:KH1}(b)] and $t_{f1}$ [see Fig.~\ref{fig:KH1}(c)]. The oscillations in the hybridization and the charge density exhibit very similar spatial patterns that are nearly isotropic in space and decay exponentially with distance from the Kondo hole. This exponential decay arises from the fact that a Kondo hole induces a
localized state outside the conduction band \cite{Fig11}. The origin of these spatial fluctuations is revealed by their wavelength $\lambda_F^c/2$, where $\lambda^c_F$ is the Fermi wave-length of the unhybridized conduction band [see Fig.~\ref{fig:KH1}(d)], implying that they arise from $2k_F^c$ scattering across the Fermi surface of the unhybridized conduction band. This result is quite unexpected, since the actual Fermi surface of the hybridized system in the heavy Fermi liquid state is large, as shown in Fig.~\ref{fig:KH1}(d). However, a strong feedback effect between the conduction electron charge density (whose spatial oscillations for sufficiently small hybridization are still determined by the unhybridized conduction electron Fermi surface) and the hybridization ensures that the hybridization oscillations exhibit a wavelength of $\lambda_F^c/2$. In contrast, the spatial oscillations of $\Delta t_{f1}$ shown in Fig.~\ref{fig:KH1}(c) extend predominantly along the lattice diagonal with a wavelength of $\lambda^h_F/2= \sqrt{2} a_0$, where $\lambda^h_F$ is the Fermi wavelength of the hybridized Fermi surface along the diagonal [see Fig.~\ref{fig:KH1}(d)]. The spatial oscillations in $t_{f1}$ therefore arise from $2k_F^h$ scattering across the Fermi surface of the hybridized bands, and their spatial form is driven by the Fermi surface's strong anisotropy [Fig.~\ref{fig:KH1}(d)] which possesses a large degree of nesting along the diagonal direction. Weaker reflections of these anisotropic oscillations can also be found in $\Delta s$, clearly demonstrating the coupling between the system's electronic and magnetic degrees of freedom.

\begin{figure}[h]
\includegraphics[height=5.0cm]{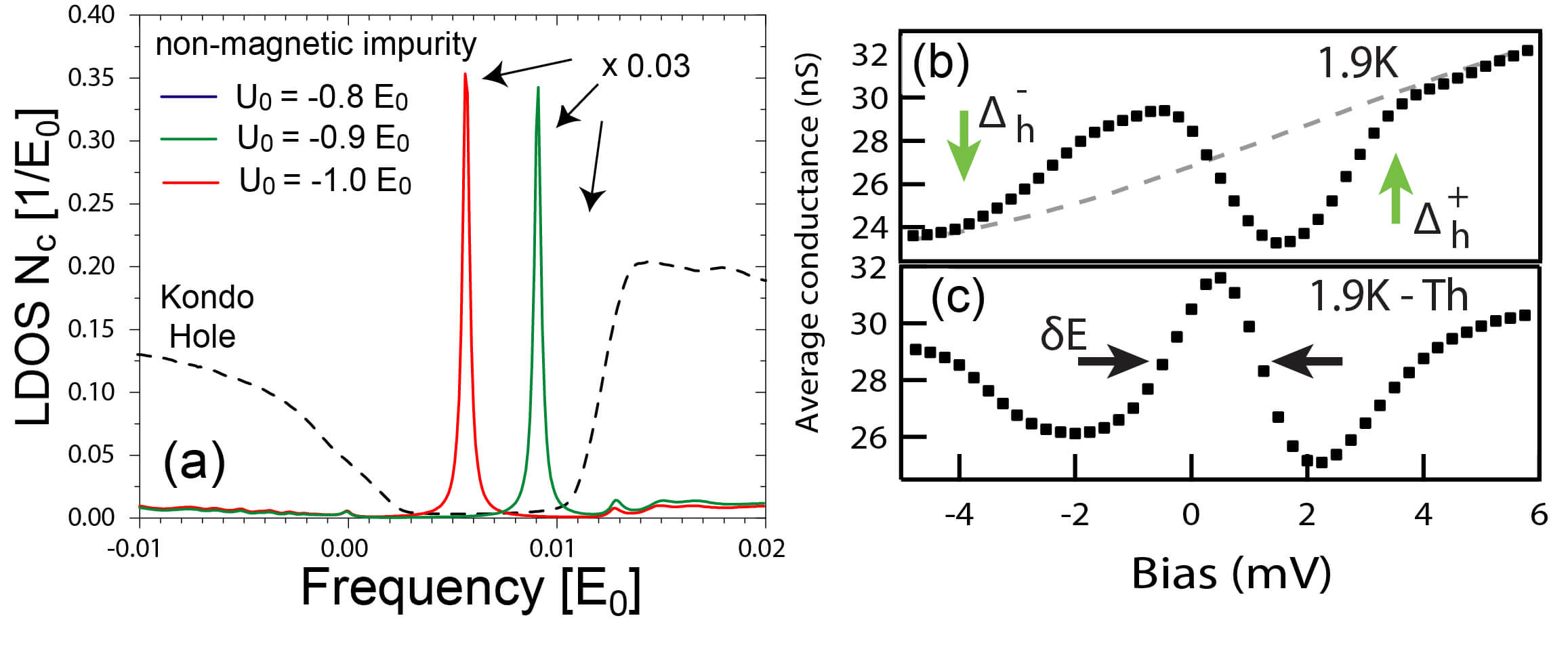}
\caption{Density of states of the conduction electrons, $N_c({\bf
R},\omega)$ for different scattering strength of a non-magnetic defect, showing the existence of impurity bound states inside the hybridization gap.  Experimental $dI/dV$ at $T=2$K (b) far from a defect, and (c) at the site of a Th atom in URu$_2$Si$_2$ \cite{Ham11}.}
\label{fig:URuSi_defect}
\end{figure}

Hamidian {\it et al.} \cite{Ham11} recently investigated the effects of defects on the local electronic structure in Th-doped URu$_2$Si$_2$. To this end, they extracted the spatial variations of the direct hybridization gap, $\Delta_h=\Delta^+_h-\Delta^-_h$, from the width of the Kondo resonance observed in $dI/dV$ [see Fig.~\ref{fig:URuSi_defect}(b)]. As the hybridization gap is twice the hybridization [see Eq.(\ref{eq:dispersions})], Hamidian {\it et al.} \cite{Ham11} were able to create a hybridization gap map that provided direct insight into the spatial variations of the hybridization induced by defects. By Fourier transforming the hybridization gap map into momentum space [see Fig.\ref{fig:KH1}(e)], they identified the characteristic wave-vector of the hybridization oscillations as twice the Fermi wave-vector of the unhybridized conduction band. This result confirms the theoretical predictions by Figgins {\it et al.} \cite{Fig11} not only of defect-induced hybridization waves in real space, but also of their characteristic wave-length [see Fig.~\ref{fig:KH1}(a)] governed by the unhybridized conduction band.

Moreover, Figgins {\it et al.}~\cite{Fig11} showed that when a magnetic moment is replaced by a non-magnetic atom that induces scattering in the conduction electron band, an impurity bound state can emerge inside the hybridization gap for sufficiently large attractive scattering potential $U_c<0$, in contrast to the effect of a Kondo hole [see dashed black line in Fig.~\ref{fig:URuSi_defect}(a)]. Its spectroscopic signature is a sharp peak in $N_c({\bf r},\omega)$ inside the hybridization gap [see Fig.~\ref{fig:URuSi_defect}(a)], that first emerges at the high energy side of the hybridization gap and then moves to lower energies with increasing $|U_c|$. The induced bound state is spatially isotropic and decays exponentially with distance from
the impurity with a decay length $\xi_D$ smaller than a lattice constant. This
small value of $\xi_D$ implies that the bound state is predominantly formed by $f$-electron states, as an impurity state formed by the light conduction electron states would possess a decay length a hundred times larger than the one observed. This result directly reflects the
strong correlations between the light and heavy bands as the defect scatters only conduction electrons, but creates a bound state that predominantly consists of $f$-electron states. The predicted impurity states inside the hybridization was subsequently observed by Hamidian {\it et al.}\cite{Ham11} in Th-doped URu$_2$Si$_2$. By comparing the differential conductance far away from a Th atom [Fig.~\ref{fig:URuSi_defect}(b)] with that at a Th atom site [Fig.~\ref{fig:URuSi_defect}(c)], they concluded that the Th atom gives rise to the emergence of an impurity state inside the hybridization gap.

\begin{figure}[h]
\includegraphics[height=4.0cm]{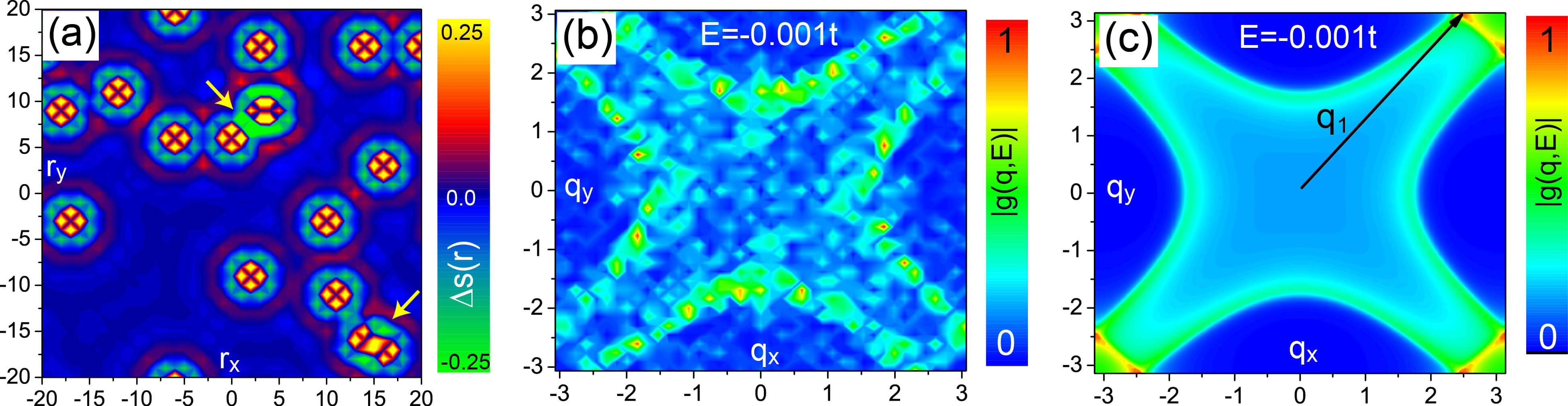}
\caption{(a) Spatial variations in the hybridization, $\Delta s({\bf r})$, arising from $1\%$ of defects in the heavy Fermi liquid state \cite{Par13}. Contour plot of the QPI intensity (b) for $1\%$ of defects, obtained from a self-consistent calculations of the electronic structure, and (c) for a single defect, obtained using the Born approximation.}
\label{fig:URuSi_disoder}
\end{figure}

Hamidian {\it et al.}\cite{Ham11} further observed that defects and disorder exert a strong effect on the electronic structure of the heavy Fermi liquid state, as a small concentration of $~1\%$ of Th atoms in URu$_2$Si$_2$ essentially disorders the entire electronic structure of the material. Using the theoretical formalism outlined in Secs.~\ref{sec:AndMod1}, Parisen~Toldin {\it et al.} \cite{Par13} found that disorder effects are enhanced in the heavy Fermi liquid state due to a strong feedback effect between the conduction electron charge density, and the hybridization, such that already an impurity concentration of $1\%$ essentially disorders the hybridization in the entire system, as shown in Fig.~\ref{fig:URuSi_disoder}(a).
Parisen~Toldin {\it et al.}  showed that while this impurity concentration possess a pronounced effect on the electronic structure as observed in the differential conductance, its effects on the material's thermodynamic properties, such as the specific heat, are rather weak (and on the order of a few percent) in agreement with the experimentally observed changes in the specific heat of heavy fermion materials with defect concentration of 1\%~\cite{Pikul}. This result explains the apparent contradiction between spectroscopic and thermodynamic measurements. Moreover, Parisen~Toldin {\it et al.} \cite{Par13} compared the QPI intensity obtained from different theoretical approaches. In particular, they computed the QPI intensity, $g({\bf q},E)$, for a heavy Fermi liquid state with $1\%$ of defects, whose effects on the electronic structure were self-consistently calculated [Fig.~\ref{fig:URuSi_disoder}(b)]. A comparison with the QPI spectrum for a single defect, obtain within the (non-self-consistent) Born approximation showed that both approaches yield the same information regarding the allowed scattering vectors, albeit with a redistribution of their spectral weight.

\section{Conclusions and Outlook}
\label{sec:Concl}

The simultaneous development of scanning tunneling spectroscopy on heavy fermion materials, together with the theoretical framework to describe and analyse it, has provided unprecedented insight into the complex electron structure of heavy fermion materials and Kondo systems. Starting from single magnetic atoms on metallic surfaces, it was possible to demonstrate that quantum interference is a crucial element in understanding how the Kondo resonance observed in the differential conductance, is related to the changes in the local electronic structure arising from the Kondo screening process and the resulting hybridization between the conduction band and the localized magnetic moments. These advances might even provide insight \cite{Yee13,Ros14} into the proposed topological nature \cite{Dze10} of the Kondo insulator SmB$_6$. Moreover, quasi-particle interference spectroscopy has revealed the momentum structure of the complex hybridized bands with unprecedented energy resolution, providing us with the opportunity to test theoretical models in great detail against experimental data. The extension of QPI spectroscopy to the superconducting state of CeCoIn$_5$ \cite{All13,Zhou13} has not only provided unique insight into the detailed momentum structure and symmetry of the superconducting order parameter, but also allowed a quantitative test of a 30-year old hypothesis for the mechanism underlying unconventional superconductivity in heavy fermion materials \cite{Dyke14}. Conducting STS experiments in the superconducting state of other heavy fermion compounds, such as recent experiments by Enayat {\it et al.}\cite{Ena15} on CeCu$_2$Si$_2$, will enable us to investigate the universality of the pairing mechanism across different families of heavy fermion materials. Equally important is the success in measuring the differential conductance in YbRh$_2$Si$_2$ by Ernst {\it et al.}\cite{Ern11} which has opened the path to investigating how the electronic structure of heavy fermion materials evolves across their quantum critical points. This, in turn, might provide the missing Rosetta Stone for understanding the emergence of non-Fermi liquid behavior in the quantum critical region. All of these advances have clearly paved the way for even more exciting discoveries in the future.

\section{ Acknowledgements}

We would like to thank M. Allan, P. Aynajian, J.C. Davis, M. Hamidian, J. Hoffman, V. Madhavan, F. Massee, H. Manoharan, J. Van Dyke, A. Yazdani, and B. Zhou for
stimulating discussions and comments.  DKM gratefully acknowledges support from the U. S. Department of Energy,
Office of Science, Basic Energy Sciences, under Award No. DE-FG02-05ER46225.

\end{document}